\documentclass[conference]{IEEEtran}

\usepackage{algorithm}

\usepackage[shortlabels]{enumitem}
\usepackage{balance}
\usepackage{booktabs}
\usepackage{todonotes}
\usepackage{epsfig}
\usepackage{graphicx}
\usepackage{amsmath}
\usepackage{amssymb}
\usepackage{bm}
\usepackage{lipsum}
\usepackage{subfig}
\usepackage{color}

\usepackage[normalem]{ulem}
\usepackage{bigstrut}
\usepackage{listings}
\usepackage{eso-pic}
\usepackage{tikz}
\usepackage{xspace}
\usepackage{fancyhdr}
\usepackage{mathtools}
\usepackage{soul}
\usepackage{tipa}
\usepackage{textgreek}
\usepackage{siunitx}
\usepackage{esvect}
\usepackage{wrapfig}
\usepackage[nocompress]{cite}

\usepackage[bookmarks=true,breaklinks=true,letterpaper=true,colorlinks,linkcolor=blue,citecolor=magenta,urlcolor=blue]{hyperref}

\def\BibTeX{{\rm B\kern-.05em{\sc i\kern-.025em b}\kern-.08em
    T\kern-.1667em\lower.7ex\hbox{E}\kern-.125emX}}

\title{\Huge Cicero: Addressing Algorithmic and Architectural Bottlenecks in Neural Rendering by Radiance Warping and Memory Optimizations}

\makeatletter
\newcommand{\linebreakand}{%
  \end{@IEEEauthorhalign}
  \hfill\mbox{}\par
  \mbox{}\hfill\begin{@IEEEauthorhalign}
}
\makeatother

\author{\IEEEauthorblockN{Yu Feng\textsuperscript{*}}
\IEEEauthorblockA{Shanghai Jiao Tong University\\University of Rochester\\y-feng@sjtu.edu.cn}
\and
\IEEEauthorblockN{Zihan Liu}
\IEEEauthorblockA{Shanghai Jiao Tong University\\altair.liu@sjtu.edu.cn}
\and
\IEEEauthorblockN{Jingwen Leng}
\IEEEauthorblockA{Shanghai Jiao Tong University\\leng-jw@cs.sjtu.edu.cn}
\linebreakand
\IEEEauthorblockN{Minyi Guo}
\IEEEauthorblockA{Shanghai Jiao Tong University\\guo-my@cs.sjtu.edu.cn}
\and
\IEEEauthorblockN{Yuhao Zhu}
\IEEEauthorblockA{University of Rochester\\yzhu@rochester.edu}
}

\newcommand*\circled[2]{\tikz[baseline=(char.base)]{
            \node[shape=circle,fill=black,inner sep=1pt] (char) {\textcolor{#1}{{\footnotesize #2}}};}}

\ifx\figurename\undefined \def\figurename{Figure}\fi
\renewcommand{\figurename}{Fig.}
\renewcommand{\paragraph}[1]{\textbf{#1} }

\newcommand{\Sect}[1]{Sec.~\ref{#1}}
\newcommand{\Fig}[1]{Fig.~\ref{#1}}

\newcommand{\proj}{\textsc{Cicero}\xspace}
\newcommand{\algo}{\textsc{SpaRW}\xspace}
\newcommand{\mode}[1]{\underline{\textsc{#1}}\xspace}

\newcommand{\no}[1]{#1}
\renewcommand{\no}[1]{#1}
\renewcommand{\hl}[1]{#1}
\newcommand{\RNum}[1]{\uppercase\expandafter{\romannumeral #1\relax}}

\def\cG{{\mathcal{G}}}
\def\cF{{\mathcal{F}}}
\def\cI{{\mathcal{I}}}

\graphicspath{{figs/}}

\begin{document}

\IEEEoverridecommandlockouts
\IEEEaftertitletext{\vspace{-1\baselineskip}}

\maketitle
\thispagestyle{plain}
\pagestyle{plain}

\begingroup\renewcommand\thefootnote{*}
\footnotetext{Work done while at University of Rochester.}
\endgroup


\setcounter{page}{1}
\begin{abstract}

Neural Radiance Field (NeRF) is widely seen as an alternative to traditional physically-based rendering. 
However, NeRF has not yet seen its adoption in resource-limited mobile systems such as Virtual and Augmented Reality (VR/AR), because it is simply extremely slow.
On a mobile Volta GPU, even the state-of-the-art NeRF models generally execute only at 0.8 FPS.
We show that the main performance bottlenecks are both algorithmic and architectural.
We introduce, \proj, to tame both forms of inefficiencies.
We first introduce two algorithms, one fundamentally reduces the amount of work any NeRF model has to execute, and the other eliminates irregular DRAM accesses.
We then describe an on-chip data layout strategy that eliminates SRAM bank conflicts.
A pure software implementation of \proj offers an 8.0$\times$ speed-up and 7.9$\times$ energy saving over a mobile Volta GPU.
When compared to a baseline with a dedicated DNN accelerator, our speed-up and energy reduction increase to 28.2$\times$ and 37.8$\times$, respectively --- all with minimal quality loss (less than 1.0~dB peak signal-to-noise ratio reduction).

\end{abstract}

\begin{IEEEkeywords}
Mobile Architecture; NeRF Acceleration; VR
\end{IEEEkeywords}

\section{Introduction}
\label{sec:intro}

Neural Radiance Field (NeRF)~\cite{mildenhall2021nerf} revives classic image-based rendering~\cite{shum2008image, szeliski2022image} using modern deep learning techniques, and has become an attractive alternative to conventional photorealistic rendering methods such as path tracing~\cite{pharr2023physically, deng2017toward, pantaleoni2010hlbvh}.
However, NeRF is excessively slow~\cite{mildenhall2021nerf}.
Despite numerous efforts~\cite{chen2022tensorf, sun2022direct, olszewski2023hashcc, yu2021plenoctrees, muller2022instant, hu2022efficientnerf, liang2022coordx}, NeRF rendering performance is still far from real-time on mobile devices.
On a mobile Volta GPU on Nvidia's Xavier SoC, common models like DirectVoxGO~\cite{sun2022direct} merely achieve 0.8 Frame Per Second (FPS), while Instant-NGP~\cite{muller2022instant} takes over 6~\si{\second} to render a $800 \times 800$ frame.

Accelerating NeRF models is critical, but one must not over-specialize for a specific model because of the rapid evolution in algorithm design.
Since its inception, NeRF models have gone through several major architectural changes, from the original grid/voxel-based design~\cite{sun2022direct} to those using hierarchical data structures ~\cite{yu2021plenoctrees, muller2022instant} and, more recently, point-based rendering through Gaussian splatting~\cite{kerbl20233d, wu2024recent, chen2024survey}.
It is conceivable that NeRF models still have to go through a few more iterations to mature.
Thus, any system and architecture support must address fundamental bottlenecks that are central to neural rendering rather than artifacts of specific models.

We identify inherent bottlenecks of NeRF models both in algorithm design and in the underlying hardware (\Sect{sec:bck}).
Algorithmically, a NeRF model has to compute the radiance of millions of rays, each carrying hundreds of ray samples.
Each ray sample executes a Multilayer Perceptron (MLP) inference, collectively incurring a high computational cost~\cite{mildenhall2021nerf}.
Architecturally, the actual computation of the MLPs introduces irregular memory accesses, stemming from the way that ray samples are grouped and accessed in NeRF.
Consequently, NeRF rendering results in many irregular DRAM accesses and SRAM bank conflicts.

This paper proposes \proj, an algorithm-architecture co-designed approach to tame both forms of inefficiency.
Algorithm-wise, we introduce a plug-and-play extension to existing NeRF algorithms, fundamentally reducing the computational workload of a NeRF model (\Sect{sec:algo}).
In particular, we exploit radiance proximity: the radiances of nearby rays emanating from the same physical point (radiant exitance) are approximately the same.
We propose \textit{sparse radiance warping} (\algo), which avoids up to 88\% of the radiance computation by reusing ray radiances computed in previous frames.
Critically, \algo is not a new NeRF model.
Rather, it is an extension that can be easily integrated into virtually all existing NeRF methods, widening its applicability.


While \algo avoids a large portion of radiance computation, \algo does not completely eliminate it.
Whenever radiance computation is needed, it is bottlenecked by irregular DRAM accesses and frequently on-chip SRAM bank conflicts.
We further propose two optimization techniques that mitigate these memory inefficiencies in radiance computation.

To eliminate irregular DRAM accesses (\Sect{sec:mem:fs}), our idea is to convert NeRF inference from a \textit{pixel-centric} order to a \textit{memory-centric} order.
The pixel-centric rendering follows the order of rays (ultimately image pixels) and their samples, resulting in discontinuous memory accesses.
The memory-centric order, in contrast, accesses voxels in the scene sequentially, inherently yielding full-streaming memory accesses.

\begin{figure*}[t]
\centering
\includegraphics[width=\textwidth]{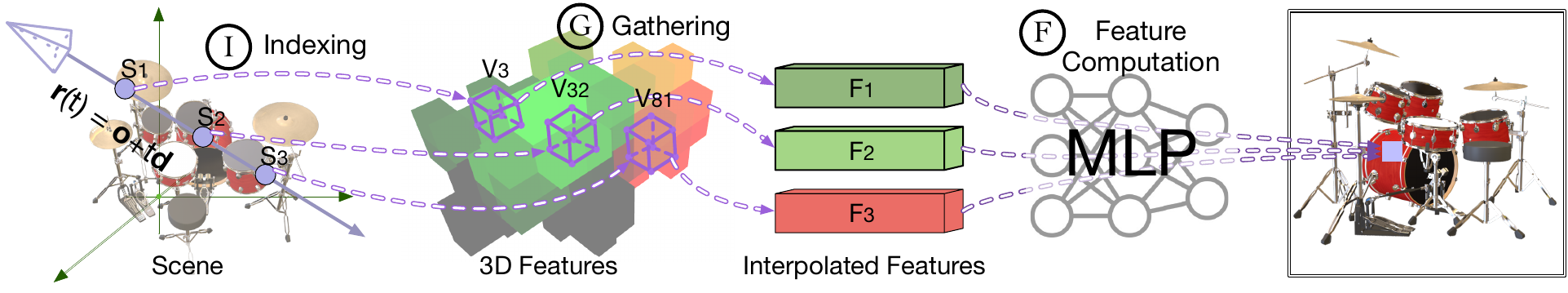}
\caption{The rendering pipeline of today's NeRF algorithms. The computation flow is highlighted in purple. Each ray first samples points, $S_{1}$, $S_{2}$, and $S_{3}$, along the ray direction. Each ray sample gathers and interpolates 3D features from eight vertices of the intersected voxel ($V_3$, and $V_{32}$, and $V_{81}$). The interpolated features ($F_1$, $F_2$, and $F_3$) are then fed into the MLP to get the partial pixel values at the three ray samples. The final pixel value is accumulated from all partial pixel values~\cite{mildenhall2021nerf}.}
\label{fig:algo_flow}
\vspace{-5pt}
\end{figure*}

To eliminate SRAM bank conflicts (\Sect{sec:mem:bank}), we explore a new data layout strategy in on-chip SRAMs.
We show that the traditional feature-major data layout, which allocates all the channels of a feature vector in the same SRAM bank, necessarily introduces frequent SRAM bank conflicts.
Instead, we propose a channel-major layout, where different channels of the same feature vector are spread across SRAM banks.
With the co-designed hardware architecture, the new data layout can completely eliminate on-chip bank conflicts.

We integrate \proj with three state-of-the-art NeRF models.
A pure software implementation achieves an 8.0$\times$ speed-up and 7.9$\times$ energy saving over a mobile Volta GPU.
Furthermore, compared to a baseline with a dedicated DNN accelerator, our speed-up and energy reduction increase to 28.2$\times$ and 37.8$\times$, respectively.

The contributions of this paper are as follows. 
\begin{itemize}
    \item We introduce \algo, a novel algorithm that reduces up to 88\% of the MLP computations in NeRF by exploiting radiance similarities between nearby rays.
    \item We propose a fully-streaming NeRF rendering algorithm that reduces the redundant DRAM access and ensures completely streaming DRAM accesses.
    \item We propose an on-chip data layout and the associated hardware support that eliminate SRAM bank conflicts.
    \item We demonstrate that \proj achieves a 28.2$\times$ speed-up and 37.8$\times$ energy savings over a baseline that has a dedicated DNN accelerator while maintaining less than 1.0~dB degradation in Peak Signal-to-Noise Ratio (PSNR).
\end{itemize}

\section{Motivation}
\label{sec:bck}

\no{We begin by discussing NeRF vs. conventional rendering methods (\Sect{sec:back:basics}).  We then overview the general pipeline of today's NeRF rendering model (\Sect{sec:back:pipeline}). We then characterize NeRF algorithms to identify Feature Gathering as a performance bottleneck (\Sect{sec:bck:comp}).
Finally, we characterize the memory inefficiencies in Feature Gathering (\Sect{sec:bck:mem}).}

\subsection{Why NeRF?}
\label{sec:back:basics}

NeRF vs. traditional ray tracing (physically-based rendering)~\cite{pharr2023physically} is widely debated in graphics.
Our work does not aim to settle that debate but, rather, to contribute to that comparison by allowing NeRF to be a more appealing option.

While NeRF is generally slower
, it has two advantages compared to ray tracing.
First, NeRF promises better rendering quality, because the complicated 
light-matter interactions are learned through data using deep learning methods rather than physically simulated.
Second, ray tracing requires a more complicated setup, e.g., modeling the geometry of the scene and describing material properties.
NeRF, in contrast, is a form of classic image-based rendering~\cite{shum2008image, szeliski2022image}, which uses a set of offline-captured images of the scene without any modeling.

From offline captured images of a scene, NeRF trains a differentiable model, which encodes the volume density and the light field in the scene (i.e., radiance of any ray)~\cite{levoy2023light, gortler2023lumigraph}.
At rendering time, given the camera pose where an image is to be rendered, the model is probed through the classic volume rendering~\cite{kaufman1993volume, levoy1988display} to synthesize the image.


\subsection{NeRF Rendering Pipeline}
\label{sec:back:pipeline}

We focus on MLP-based models.
More recent 3D Gaussian Splatting (3DGS) models~\cite{kerbl20233d, wu2024recent, chen2024survey} do away with MLPs; they are generally faster (still far from real-time on mobile devices) at the cost of much larger model sizes.
How \proj can be applied to 3DGS models will be discussed in \Sect{sec:lim}.
State-of-the-art NeRF models have a general pipeline that consists of three stages: Indexing ($\cI$), Feature Gathering ($\cG$), and Feature Computation ($\cF$), as illustrated in \Fig{fig:algo_flow}.

\paragraph{Indexing ($\cI$).}
The entire 3D scene is initially partitioned into many adjacent voxels.
Each voxel is a cube with eight vertices in the space.
Each vertex carries a high-dimensional feature vector that is offline trained.
During rendering, we first generate a ray for each pixel to be rendered.
Each ray samples a set of points (e.g. $S_1$, $S_2$, and $S_3$ in \Fig{fig:algo_flow}) along its direction.
Each ray sample, based on its sampled position in the space, calculates the ID of the voxel that contains the sample.

\paragraph{Feature Gathering ($\cG$).} 
Using the voxel ID, each ray sample finds the eight vertices of that voxel and gathers the features of the eight vertices.
In the example in \Fig{fig:algo_flow}, $S_1$ would access the eight vertex features in $V_3$. 
These features encode both the density and radiance field at the corresponding vertex locations. 
This ray sample then computes its feature by trilinearly interpolating the feature vectors of the eight vertices. 


\paragraph{Feature Computation ($\cF$).}
In Feature Computation, each ray sample passes its intermediate feature through a lightweight MLP model to obtain the actual density and radiance value of that sample.
The final color of a ray (and thus the color of the pixel that is hit by the ray) would be computed by accumulating all ray samples along the ray direction.



\subsection{Computation Characterizations}
\label{sec:bck:comp}

\begin{figure}[t]
\centering
\begin{minipage}[t]{0.48\columnwidth}
  \centering
  \includegraphics[width=\columnwidth]{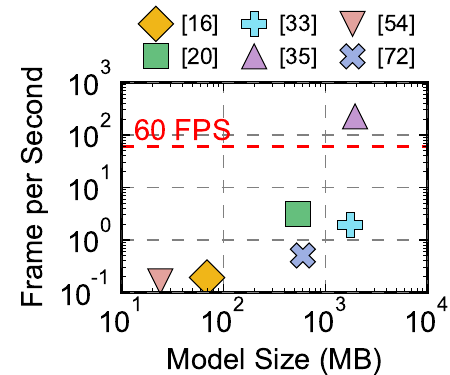}
  \caption{Frame rate vs. model size on the Xavier SoC~\cite{xaviersoc}. Models~\cite{muller2022instant, sun2022direct, chen2022tensorf, hu2022efficientnerf, chen2023mobilenerf, hedman2021baking} are named by reference numbers.}
  \label{fig:compute_vs_model_size}
\end{minipage}
\hspace{2pt}
\begin{minipage}[t]{0.48\columnwidth}
  \centering
  \includegraphics[width=\columnwidth]{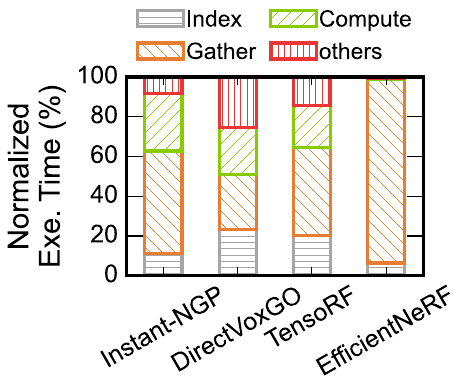}
  \caption{Normalized execution breakdown across state-of-the-art NeRF algorithms~\cite{muller2022instant, chen2022tensorf, hu2022efficientnerf, sun2022direct}.}
  \label{fig:execution_dist}
\end{minipage}

\end{figure}

\paragraph{Performance and Model Size.}
Today's NeRF models are not only slow, but they also have model sizes that do not fit in on-chip SRAMs.
\Fig{fig:compute_vs_model_size} shows the frame rate on a mobile Volta GPU~\cite{xaviersoc} ($y$-axis) and model size ($x$-axis) of several state-of-the-art NeRF models~\cite{muller2022instant, sun2022direct, chen2022tensorf, hu2022efficientnerf, chen2023mobilenerf, hedman2021baking}.
We also overlay the 60 FPS frame rate requirement.
NeRF algorithms rarely achieve real-time rendering on today's commodity mobile devices.

In addition, NeRF model sizes far exceed the on-chip SRAM sizes affordable on today's mobile SoC, necessitating frequent DRAM accesses.
A NeRF model's size includes both the feature vectors of the voxels and the MLP model weights; the former dominates the total size.
The feature vectors are usually at the order of 10 MB -- 1,000 MB, whereas the MLP weights are generally small (10 KB -- 100 KB).

\paragraph{Performance Bottleneck.}
We characterize the performance bottlenecks across NeRF algorithms. 
\Fig{fig:execution_dist} shows the execution breakdown of different stages across four popular NeRF algorithms on a mobile Volta GPU~\cite{xaviersoc}. All three stages take non-trivial execution time with Feature Gathering dominating the execution ($>$56\% of execution time on average).

\subsection{Memory Inefficiencies in NeRF}
\label{sec:bck:mem}

Given that Feature Gathering is memory-heavy, we further characterize its memory accesses.
We find that, while Feature Gathering is computationally parallel (between rays and between samples on a ray), it is not memory-friendly, introducing both irregular DRAM accesses and frequent on-chip bank conflicts.
The following paragraphs provide a quantitative analysis to illustrate these issues.

\begin{figure}[t]
\centering
\begin{minipage}[t]{0.48\columnwidth}
  \centering
  \includegraphics[width=\columnwidth]{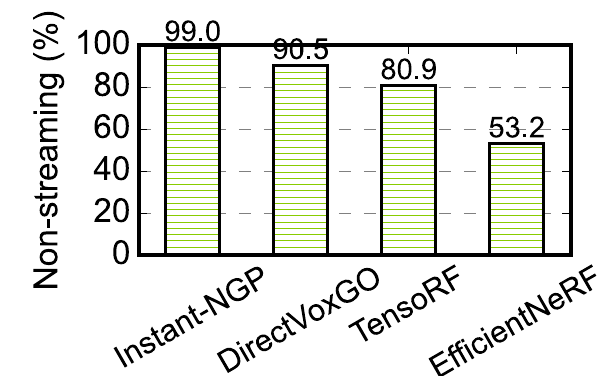}
  \caption{Percentage of non-continuous DRAM accesses in feature gathering.}
  \label{fig:dram_non_stream}
\end{minipage}
\hspace{2pt}
\begin{minipage}[t]{0.48\columnwidth}
  \centering
  \includegraphics[width=\columnwidth]{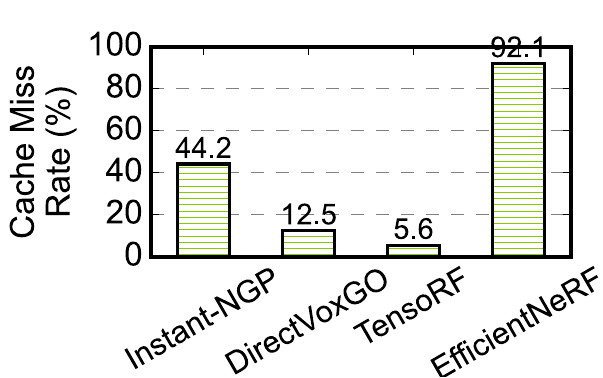}
  \caption{Cache miss rate in feature gathering across common NeRF algorithms.}
  \label{fig:cache_miss}
\end{minipage}
\end{figure}

\paragraph{DRAM Access Inefficiency.} The inefficiency in DRAM access can be attributed to two main factors: non-streaming DRAM access and redundant DRAM access, both stemming from the inherent \textit{pixel-centric} rendering in NeRF models.

NeRF inference is parallelized across pixels, where adjacent pixels in the final image are computed simultaneously.
This pixel-centric rendering introduces two levels of memory-access irregularities: inter-ray irregularity and intra-ray irregularity. 
Inter-ray irregularity means that rays of adjacent pixels might access non-continuous memory regions.
This is because rays might diverge as they transport over space, even when their origins are spatially close.


Intra-ray irregularity is caused by sampling points along a single camera ray accessing discontinuous memory regions.
Specifically, the feature vectors corresponding to the different ray samples can be stored at arbitrary memory locations.
As shown in \Fig{fig:algo_flow}, ray samples, $S_1$ and $S_2$, intersect voxels, $V_3$ and $V_{32}$, which are spatially distant and, thus, not stored continuously in DRAM.
\Fig{fig:dram_non_stream} shows the non-streaming DRAM access in four popular NeRF algorithms.
On average, over 81\% of DRAM access is non-streaming.

Irregular accesses mean that each voxel might be accessed multiple times during rendering, which leads to redundant DRAM accesses, given that the feature vectors cannot be stored completely on-chip (\Fig{fig:compute_vs_model_size}).
Assuming a 2 MB on-chip buffer with oracle replacement~\cite{xaviersoc}, \Fig{fig:cache_miss} shows the cache miss rates of different NeRF algorithms during feature gathering; the miss rate can be as high as 92\% (average 38\%).
In reality, an even smaller buffer would be allocated to accommodate other operations and data structures.

\begin{figure}[t]
\centering

\begin{minipage}[t]{0.48\columnwidth}
  \centering
  \includegraphics[width=\columnwidth]{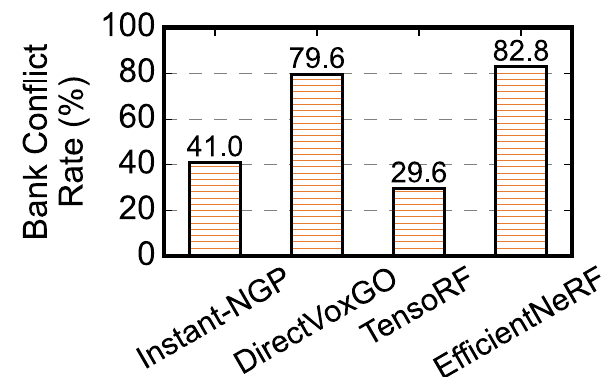}
  \caption{SRAM bank conflict rate in feature gathering, assuming 16 banks and 16 concurrent ray queries.}
  \label{fig:sram_conflict}
\end{minipage}
\hspace{2pt}
\begin{minipage}[t]{0.48\columnwidth}
  \centering
  \includegraphics[width=\columnwidth]{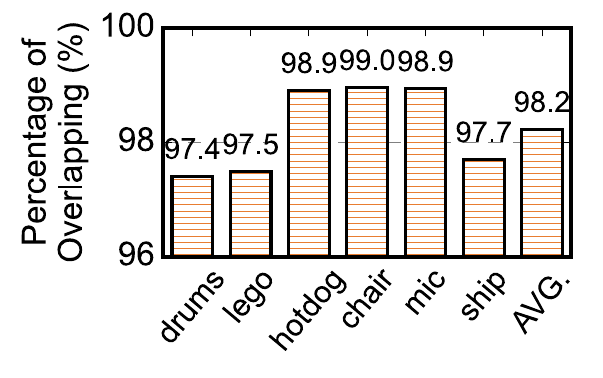}
  \caption{The overlapping percentage across six scenes in Synthetic-NeRF~\cite{mildenhall2021nerf}.}
  \label{fig:warp_pct}
\end{minipage}

\end{figure}

\paragraph{SRAM Access Inefficiency.}
On-chip memory access in NeRF results in frequent bank conflicts.
In conventional DNNs, the memory access patterns can be determined statically. Thus, bank conflicts can be eliminated through meticulous data layout across SRAM banks~\cite{zhou2021characterizing, kirk2016programming}.
Conversely, the data access pattern of Feature Gathering in NeRF depends on the camera view and cannot be known offline. 

\Sect{sec:mem:bank} will provide a detailed description of the causes of bank conflicts in Feature Gathering. Here, we simply show the bank conflict rate of Feature Gathering across NeRF algorithms (\Fig{fig:sram_conflict}). Assuming a 2 MB buffer with 16 banks and 16 concurrent camera rays, the average bank conflict rate is 52\%, with EfficientNeRF reaching as high as 83\%.
A larger number of concurrent rays would lead to a higher bank conflict rate. For instance, the bank conflict rate of Instant-NGP increases to 80\% when the number of rays escalates to 64.
Increasing the number of banks does reduce the bank conflict rate. However, heavily banked SRAM designs are highly undesirable due to costly crossbars~\cite{agarwal2009garnet, grot2011kilo}.


\section{Sparse Radiance Warping}
\label{sec:algo}

\no{This section introduces, \textit{sparse radiance warping} (\algo), an algorithm that exploits the radiance similarity across rays from nearby camera views.
We first provide an intuition of \algo algorithm (\Sect{sec:algo:intuition}), followed by a description of the overall algorithm (\Sect{sec:algo:main}).
Finally, we discuss two key aspects in the \algo algorithm design that helps improve performance and rendering quality (\Sect{sec:algo:dd}).
}


\subsection{Intuition}
\label{sec:algo:intuition}

\begin{figure}[t]
  \centering
  \includegraphics[width=0.85\columnwidth]{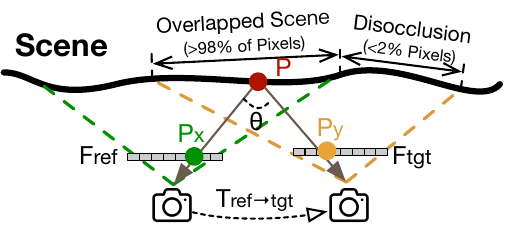}
  \caption{Intuition of our \algo algorithm. The radiance of the $\overline{P P_x}$ ray can approximate the radiance of the $\overline{P P_y}$ ray if the angle $\theta$ between these two rays is sufficiently close.}
  \label{fig:intuition}
\end{figure}

The goal of \algo is to reuse pixel values rendered in previous frames using a technique called image warping.
\Fig{fig:intuition} illustrates our idea, which starts from a previously rendered frame, called a reference frame, $F_{ref}$.
For a given pixel in $F_{ref}$, say $P_x$, we can find the point in the scene $P$ that is captured by that pixel.
When rendering a new target frame $F_{tgt}$, the same point $P$ is captured as a new pixel $P_y$ in the target frame $F_{tgt}$.
The assumption here is that if the camera poses of $F_{tgt}$ and $F_{ref}$ are sufficiently close, 
the radiance of both $\overline{P P_x}$ ray and $\overline{P P_y}$ ray are approximately similar.
Thus, the pixel value $P_x$ can be simply reused in $P_y$, avoiding rendering $P_y$ through the compute-intensive NeRF model.


This warping idea avoids rendering pixels in $F_{tgt}$ whose corresponding scene points are also captured by $F_{ref}$.
The larger the overlap between $F_{ref}$ and $F_{tgt}$ is, the less NeRF computation is required.
\Fig{fig:warp_pct} characterizes the overlapping between two adjacent frames in the Synthetic-NeRF dataset~\cite{mildenhall2021nerf}.
More than 98\% of pixels are overlapped (standard deviation: 1.7\%), indicating that less than 2\% of pixels require re-rendering.
The same conclusion holds for real-world datasets:
on the Unbounded-360~\cite{barron2022mipnerf360} and Tanks and Temples~\cite{Knapitsch2017} datasets, only 4.3\% and 4.9\% pixels cannot be warped, respectively.
The high overlap is not an artifact of a particular dataset but a fundamental attribute of real-time rendering, where consecutive frames are necessarily in close proximity because the observer/camera does not jump arbitrarily.

\begin{figure}[t]
  \centering
  \includegraphics[width=\columnwidth]{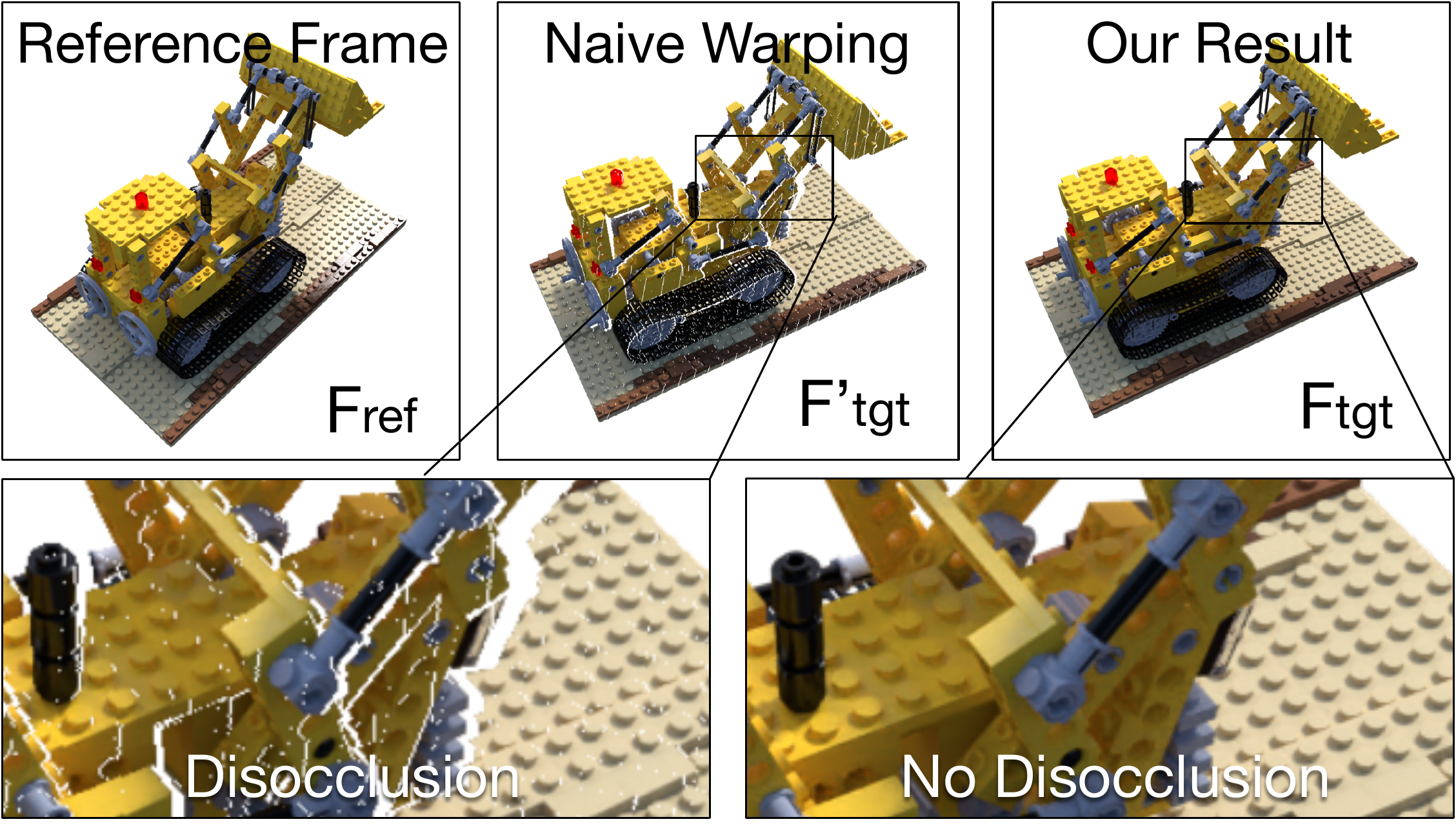}
  \caption{Examples of a reference frame $F_{ref}$, a result of naive warping $F'_{tgt}$ and our result $F_{tgt}$ by \algo. Note that disocclusions (missing pixels) are eliminated in our result.}
  \label{fig:res_demo}
\end{figure}

The non-overlapped pixels, called \textit{disoccluded pixels}, arise when the previously occluded scene in $F_{ref}$ becomes visible in $F_{tgt}$.
\Fig{fig:res_demo} shows the effect of a naive warping.
Without recalculating the disoccluded pixels, the rendered image $F'_{tgt}$ has clear ``holes'', because the disoccluded pixels cannot be warped from the reference frame.
Our idea then is to calculate the disoccluded pixels using the original NeRF model, which now renders only a small amount of (e.g., 2\%) sparsely disoccluded pixels in the target frame.

\begin{figure*}[t]
\centering
\includegraphics[width=2\columnwidth]{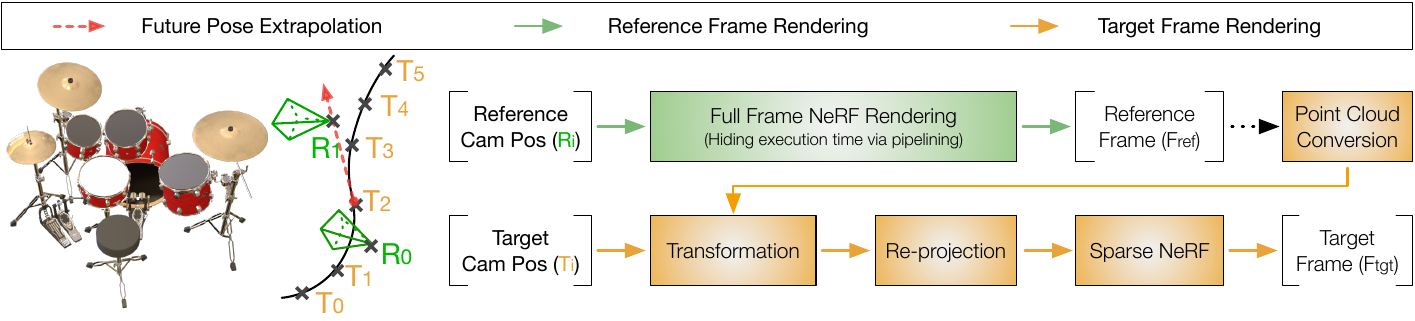}
\caption{An overview of the \algo algorithm.
Only reference frames ($R_i$) undergo full-frame NeRF inference as denoted by the green path.
All target frames ($T_i$) are computed using the lightweight warping operations denoted by the orange path.
The reference frames are not on the camera trajectory so reference frame rendering and target frame rendering can be overlapped (\Fig{fig:pipeline}).
Camera poses at reference frames are extrapolated using the poses of previous target frames.
}
\label{fig:algo}
\end{figure*}

\subsection{Basic Algorithm}
\label{sec:algo:main}

In \algo there are two rendering paths, which are illustrated in \Fig{fig:algo}: a compute-intensive path (in green) to render reference frames ($R_0$ and $R_1$) using full-frame NeRF rendering, and a lightweight path (in orange) that uses the warping idea in \Sect{sec:algo:intuition} to render target frames ($T_0$ -- $T_5$).
We first describe how to warp from a reference frame to a target frame, then discuss the choice of reference frames.

\paragraph{Target Frame Rendering.} 
There are four steps in rendering a target frame: \circled{white}{1} point cloud conversion, \circled{white}{2} transformation, \circled{white}{3} re-projection and \circled{white}{4} sparse NeRF rendering. 

\circled{white}{1} Given a reference frame $F_{ref}$, we first convert $F_{ref}$ into a point cloud $P_{ref}$, which represents the 3D scene in the reference camera coordinate system.
The transformation uses scene depth and the camera's intrinsic parameters. Mathematically, it can be expressed as follows:
\begin{gather}
    P_{ref} = \begin{bmatrix}
        \frac{D_{ref}}{f} & 0 & -\frac{D_{ref} C_x}{f} \\
        0 & \frac{D_{ref}}{f} & -\frac{D_{ref} C_y}{f} \\
        0 & 0 & D_{ref}
    \end{bmatrix} \times F_{ref}
\end{gather}
where $D_{ref}$ is the depths of points in $P_{ref}$ corresponding to pixels in $F_{ref}$; $D_{ref}$ can be obtained through a standard rasterization pipeline (using depth buffer)~\cite{shirley2009fundamentals};
$f$ is the camera focal length;
[$C_x$, $C_y$] is the camera center.
Both focal length and camera center are part of the camera's intrinsic parameters~\cite{forsyth2002computer}.

\circled{white}{2} The $P_{ref}$ calculated so far is expressed in the coordinate system of the reference frame.
To render the target frame, we must transform the point cloud to the coordinate system of the target frame --- using a simple linear transformation:
\begin{equation}
    P_{tgt} = T_{ref \rightarrow tgt} \times P_{ref}
\end{equation}
where $T_{ref \rightarrow tgt}$ is the transformation matrix between the reference camera pose $R_i$ and the target camera pose $T_i$;
$P_{tgt}$ denotes the point cloud in the target frame's coordinate system.

\circled{white}{3} Once we have $P_{tgt}$ expressed at the current camera coordinate system, obtaining the frame at the current camera pose requires a standard perspective projection in the classic rasterization pipeline~\cite{shirley2009fundamentals}:
\begin{gather}
    F'_{tgt} = \begin{bmatrix}
        \frac{f}{D_{tgt}} & 0 & 0 & C_x\\
        0 & \frac{f}{D_{tgt}} & 0 & C_y \\
        0 & 0 & \frac{1}{D_{tgt}} & 0
    \end{bmatrix}  \times P_{tgt}
\end{gather}
\noindent where $D_{tgt}$ is the depth of all the points corresponding to the pixels in the target frame.

\circled{white}{4}
As shown in \Fig{fig:res_demo}, naively warped frame $F_{tgt}'$ contains disocclusion artifacts.
To mitigate disocclusions, we simply run the original NeRF model for those disoccluded pixels,
\begin{equation}
    F_{tgt} = F'_{tgt} \circledast \Gamma_{sp}
\end{equation}
\noindent $\Gamma_{sp}$ denotes sparse NeRF rendering of disoccluded pixels, and $\circledast$ combines the warped pixels with the NeRF-rendered pixels.

Interestingly, ``holes'' in a target frame can be attributed to two factors: disocclusion and void (i.e., areas in the scene where there is nothing).
To avoid unnecessary computation on the latter, we perform a simple depth test so that pixels whose depth is infinite are skipped in sparse NeRF rendering.
The depth map of the current frame can be obtained through, again, the standard perspective projection.
The overhead of such a projection is minimal.
In our measurement, the latency of processing, e.g., one million points, is less than one millisecond on a Nvidia Volta mobile GPU.

\subsection{Key Design Decisions}
\label{sec:algo:dd}

\paragraph{Reference Frame Rendering.}
Target frame rendering relies on the existence of a reference frame, which in \algo will rendered via full-frame NeRF rendering.
The key question is \textit{how to choose the reference frames}.
Technically any frame can be a reference frame, and the reference frames do not even have to be on the camera's trajectory.
In \Fig{fig:algo}, the two reference frames $R_0$ and $R_1$ are indeed off the trajectory.

\begin{figure}[t]
\centering
\subfloat[Reference and target frame rendering must be serialized if reference frames are chosen to be on the camera trajectory. $R_1$ can be rendered only after $T_{k-1}$.]{
	\label{fig:prev_pipeline}	\includegraphics[width=0.98\columnwidth]{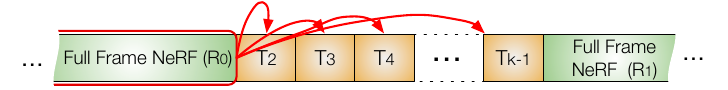} }
\\
\vspace{-5pt}
\subfloat[Reference and target frame rendering can be overlapped if reference frames are off the trajectory. $R_1$ can be rendered simultaneously with $T_2$ -- $T_{k-1}$.]{
	\label{fig:our_pipeline}
	\includegraphics[width=0.98\columnwidth]{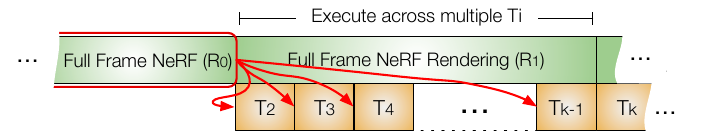} } 
\caption{Two choices of reference frames. Red arrows show how a reference frame is used to warp target frames, which is the same between the two methods. Our method (bottom) overlaps reference and target frame rendering.}
\label{fig:pipeline}
\end{figure}

Rendering reference frames on the trajectory is a common strategy in previous work that uses temporal correlation~\cite{zhu2018euphrates, feng2019asv, song2020vr, buckler2018eva2}.
The advantage is that it is work-efficient: on-trajectory frames have to be rendered anyway.
However, this approach also limits \textit{when} a reference frame can be rendered: a reference frame can only be rendered when its actual camera pose is obtained.
As a result, it necessarily serializes target frame rendering and reference frame rendering, as illustrated in \Fig{fig:prev_pipeline}, because all frames (target or reference) must be rendered in the order in which the camera poses are obtained, which is necessarily sequential.

Instead, our observation is that the reference frames do not have to be actual frames that users see; they just provide information that can be reused in the target frames.
As long as the reference frame is close to the camera trajectory, the radiance approximation still holds.
This allows us to overlap reference frame rendering with target frame rendering.

This overlapping idea is illustrated in \Fig{fig:algo} with the corresponding timeline illustrated in \Fig{fig:our_pipeline}.
The reference frame $R_1$ is off the trajectory.
The camera position of $R_1$ is chosen so that it is close to the trajectory.
In our implementation, we use the velocity at camera pose $T_2$ to extrapolate the position of $R_1$~\cite{hou2020motion}.
This does not mean that rendering $R_1$ depends on the rendering result of $T_2$; rather, it depends only on the camera pose of $T_2$, which is known before $T_2$ is rendered.

Specifically, we use the position of the last two rendered frames, $T_1$ and $T_2$, to calculate the velocity $v$ at $T_2$:
\begin{align}
v = \frac{T_2 -  T_1}{\Delta t},
\end{align}

\noindent where $\Delta t$ represents the interval between two consecutive frames.
We then calculate the pose of the reference frame $R_1$:
\begin{align}
	R_1 = T_2 + v \times t_r,~~~t_r = \frac{N}{2}\Delta t,
\end{align}

\noindent where $N$ is the number of target frames that share the same reference frame (i.e., 4 in this example as $T_2$ -- $T_5$ share $R_1$).
Using $\frac{N}{2}$ allows the reference frame to be roughly at the center of, and thus increase the overlap with, its target frames.

Critically, while rendering $R_1$ using a compute-intensive full-frame NeRF model, we can \textit{simultaneously} render the target frames $T_2$ -- $T_5$ (using the lightweight warping computation) based on a previous reference frame $R_0$, whose position itself is extrapolated from an earlier target frame such as $T_0$.
We show that one single reference can be used for up to 30 target frames with a minimal visual quality loss (\Sect{sec:eval:sen}).
In this way, the long latency of rendering reference frames can be largely hidden behind rendering target frames.

\paragraph{Deciding When to Warp.}
A potential limitation of our \algo algorithm is that the radiance approximation does not hold well when 1) consecutive frames have large differences in camera poses (i.e., $\theta$ in \mbox{\Fig{fig:intuition}} is too large) and 2) the surface material is non-diffuse.
As a simple heuristic to mitigate such issues, we allow warping only when the angle subtended by a ray in the reference frame and the corresponding ray in the target frame (i.e., $\theta$ in \Fig{fig:intuition}) is smaller than a threshold $\phi$.
We provide a more comprehensive discussion of this warping heuristics in \Sect{sec:lim}.

\begin{figure*}[t]
\centering
\includegraphics[width=2\columnwidth]{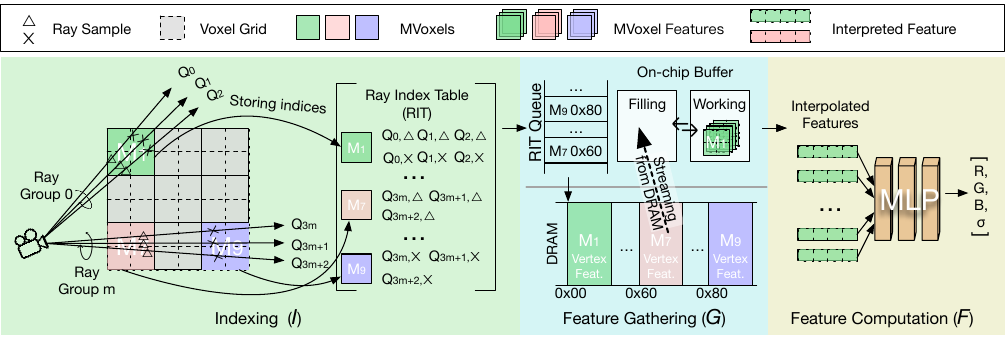}
\vspace{-5pt}
\caption{Fully-streaming NeRF rendering algorithm. We first group all the voxels into MVoxels, which are continuously stored in the DRAM.
The Ray Index Table (RIT) records, for each MVoxel, the IDs of ray samples whose features reside in the MVoxel.
During feature gathering ($\cG$), the entries in the RIT are sequentially accessed, essentially streaming the MVoxels from the DRAM.
Each time an MVoxel is loaded on-chip, we process all the ray samples whose feature vectors are in that MVoxel.
\no{The Feature Computation stage is unchanged.}
}
\vspace{-5pt}
\label{fig:streaming}
\end{figure*}




\section{Memory Optimizations}

While \algo reduces NeRF computation of target frames, reference frames still execute full-frame NeRF rendering, which is bottlenecked by redundant and irregular DRAM accesses of feature vectors and the frequent on-chip bank conflicts as shown in \Sect{sec:bck:mem}.
This section first describes an algorithmic optimization that eliminates redundant DRAM accesses and guarantees fully-streaming DRAM accesses (\Sect{sec:mem:fs}).
We then discuss a new on-chip data layout that eliminates SRAM bank conflicts (\Sect{sec:mem:bank}).
Finally, we describe our co-designed hardware architecture that unleashes the two memory optimizations (\Sect{sec:mem:hw}).

\subsection{Fully-Streaming NeRF Rendering}
\label{sec:mem:fs}




\paragraph{Architectural Assumptions.}
We assume a DNN accelerator for MLP operations.
The accelerator has an on-chip buffer (scratchpad) to store 3D voxel encoding (feature vectors) in the NeRF model.
However, this buffer is generally too small (1~MB -- 3~MB) to hold the entire 3D voxel encoding (10~MB -- 1000~MB).
Additionally, there is a dedicated on-chip buffer to store MLP weights of NeRF models; these weights are generally small (10~KB -- 100~KB).


\paragraph{Memory-Centric Rendering.}
Recall from \Sect{sec:bck:mem} that redundant and irregular DRAM accesses are caused by the \textit{pixel-centric} rendering, where we parallelize the computation across all the pixels in the order that they appear in the final image.
As a result, the ray samples during Feature Gathering ($\cG$) access discontiguous memory regions.

Instead, we propose a \textit{memory-centric} rendering, where the order of computation is based on where the ray samples reside in the DRAM.
At a high level, we sequentially read, in chunks, voxel features that are contiguously laid in memory.
As each chunk is loaded to the on-chip SRAM, we render the ray samples whose feature vectors happen to reside in the chunk.
We throw away the chunk only after all the associated ray samples have been computed.
In this way, we guarantee that each voxel feature is read only once and the DRAM accesses to the voxel features are fully streaming.

\hl{Memory-centric rendering incurs no storage overhead: the vertex features are stored in memory \textit{as is} without duplication.
What is being reordered is the order we \textit{access} the features.}
We use \Fig{fig:streaming} to illustrate the idea and how the three stages in NeRF rendering are changed accordingly.

\paragraph{Indexing ($\cI$).}
We first group all the voxel features into macro voxels (MVoxels).
All the data in a MVoxel is loaded to the SRAM together when the MVoxel is loaded.
In the example of \Fig{fig:streaming}, we combine every $2\times2$ voxels into one MVoxel.
In reality, we guarantee that the data size of one MVoxel is smaller than the on-chip buffer size.
We store vertex features within one MVoxel continuously in the DRAM, and store MVoxels continuously in the DRAM.

We then compute a Ray Index Table (RIT), where each MVoxel has an entry.
Each entry records the IDs of all the ray samples whose features reside in that particular MVoxel.
Note that the ray sample-to-voxel mapping has to be calculated in the original NeRF models too; we simply group all such calculations and store the results in a table.



\paragraph{Feature Gathering ($\cG$) and Feature Computation ($\cF$).} 
During gathering, we load the MVoxels from the DRAM to the SRAM sequentially.
When an MVoxel is loaded, we look up the RIT to find all the ray samples that can be computed.
A standard double buffer can be used to overlap MVoxel loading and the on-chip computation.
Feature Computation remains unchanged compared to the baseline NeRF rendering.





\paragraph{Accommodating Hierarchical Data Encodings.}
Some NeRF algorithms, instead of storing voxel features directly, use hierarchical data structures such as hierarchical voxel grid~\cite{muller2022instant, sun2022direct, chen2023mobilenerf}, hashing~\cite{muller2022instant, olszewski2023hashcc}, and factorized tensor~\cite{chen2022tensorf} to index the features. 

To accommodate our fully-streaming data flow with these hierarchical data structures, we first partition 3D voxels at each level into MVoxel grids.
During Feature Gathering, we group all rays into small ray groups (\Fig{fig:streaming}) and collect features level-by-level for a given ray group.
Once we have traversed all levels, we can then compile all the vertex features necessary for this ray group.
\hl{When the 3D voxel dimensions in the last several levels are too large, loading each MVoxel entirely would lead to low utilization of voxels, wasting DRAM bandwidth.
In that case, we revert back to the original (non-streaming) data flow.
This reversion happens in, for instance, Instant-NGP~\mbox{\cite{muller2022instant}} from level 5 (out of 8 levels) onwards.
As a result, about half of the DRAM traffics on Instant-NGP are non-streaming (which is faithfully captured in evaluation).}

\subsection{Bank Conflict-Free Interleaving}
\label{sec:mem:bank}

\begin{figure}[t]
\centering
\subfloat[The feature vector layout in existing NeRF accelerators, where vertex features are spread across SRAM banks, but all the channels in the same feature vector are stored in the same bank.]{
	\label{fig:bank_conflict_exp}	\includegraphics[width=0.98\columnwidth]{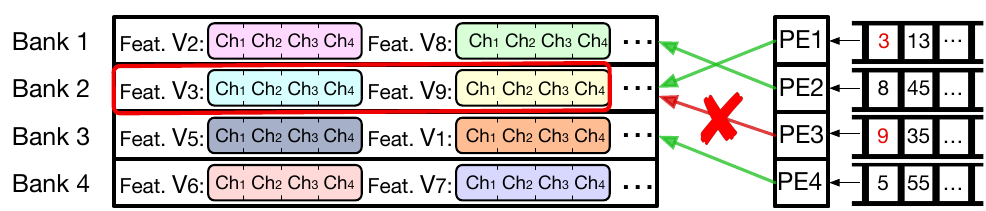} }
\\
\vspace{-5pt}
\subfloat[Our data layout spreads channels of a feature vector across different banks.]{
	\label{fig:our_layout}
	\includegraphics[width=0.98\columnwidth]{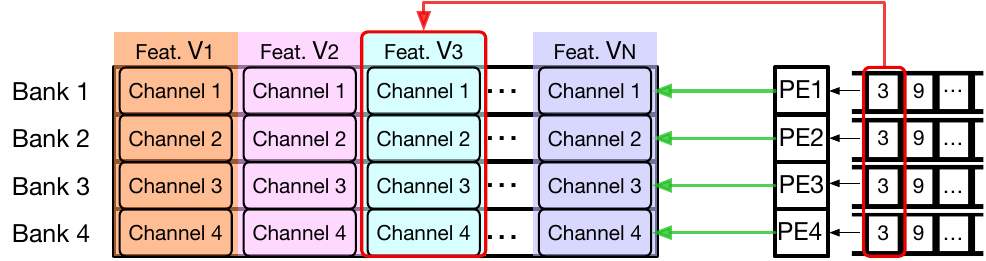} } 
\caption{A comparison between the original feature vector layout (\Fig{fig:bank_conflict_exp}) and our data layout (\Fig{fig:our_layout}).
The example has four banks and four concurrent PEs.
In \Fig{fig:bank_conflict_exp}, a bank conflict occurs when PE$_1$ and PE$_3$ (each collecting features for a different ray sample) access two different features from bank 2.
Our data layout (\Fig{fig:our_layout}) eliminates bank conflicts by 1) spreading channels across banks and 2) having each PE collect a particular channel across different ray samples.}
\label{fig:bank_conflict_comp}
\end{figure}

With fully-streaming DRAM accesses, the inefficiency shifts to the on-chip SRAM, which experiences frequent bank conflicts (\Fig{fig:sram_conflict}),
which arise when different rays access the vertex features located at the same bank, leading to stalls in Feature Gathering ($\cG$). 
Critically, unlike conventional DNNs where one can orchestrate data layout offline to avoid bank conflicts~\cite{zhou2021characterizing, kirk2016programming}, the SRAM access pattern of ray samples is known only at the run time, because the exact ray samples depend on the run-time camera pose information.


The reason behind bank conflicts in Feature Gathering has to do with the way feature vectors are laid out in SRAM; \Fig{fig:bank_conflict_exp} illustrates this point.
State-of-the-art NeRF accelerators~\cite{li2023instant, lee2023neurex} store feature vectors in the SRAM using a \textit{feature-major} order, where all the channels of a feature vector are stored in the same SRAM bank.
Assume in this example we have four PEs, each responsible for collecting the feature vector for a particular ray sample.
PE$_1$ and PE$_3$ are responsible for two ray samples, which require feature vectors 3 and 9, respectively.
However, the two feature vectors happen to reside in the same bank, causing a bank conflict.

To address this issue, we propose a \textit{channel-major} layout, as illustrated in \Fig{fig:our_layout}, where different channels of the same feature vector are spread across banks.
For instance, bank 1 stores the first channel of all feature vectors within one MVoxel.
In cases where the feature channel size exceeds the bank size, the storing sequence restarts from bank 1.

During feature gathering, instead of parallelizing ray samples across PEs, we parallelize channels across PEs.
Each PE is responsible for gathering one channel of all the ray samples rather than gathering all channels of one individual ray sample.
That is, each PE is dedicated to a specific bank.
For instance, in \Fig{fig:our_pipeline}, the four PEs are collecting the four channels of the same feature vector 3 (required by one ray sample).

\begin{figure}[t]
  \centering
  \includegraphics[width=\columnwidth]{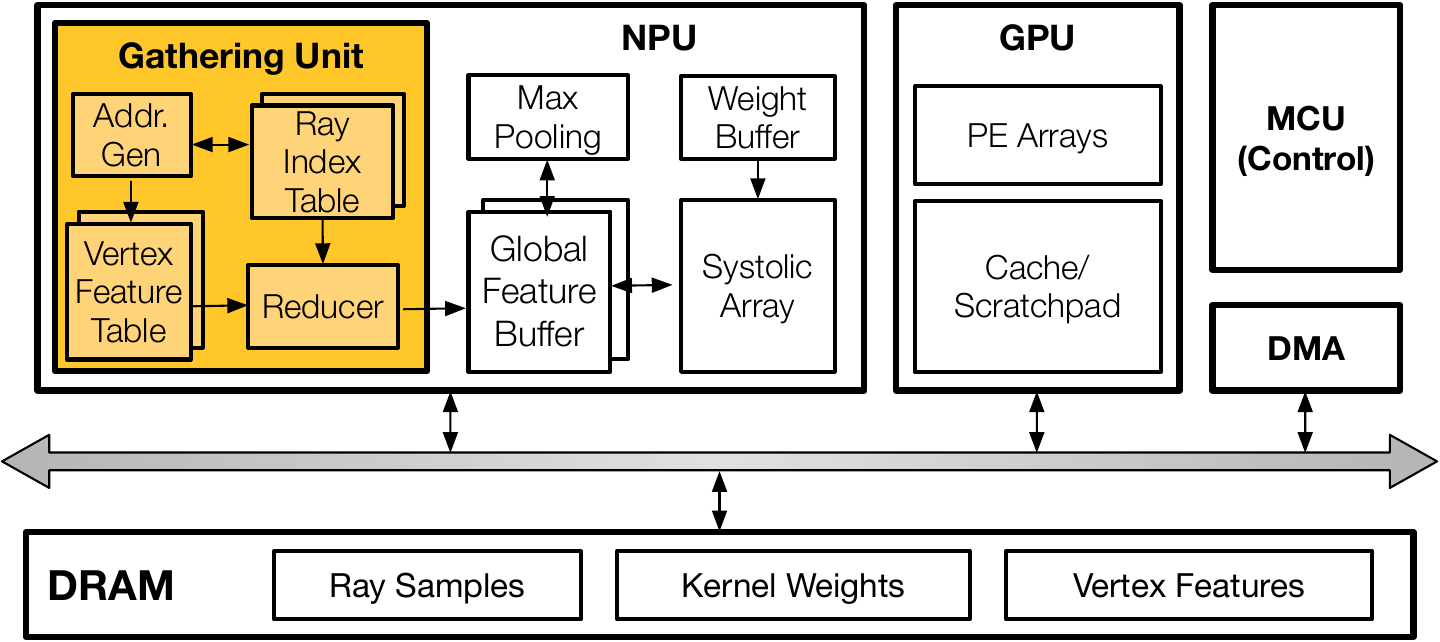}
  \caption{The SoC architecture; the uncolored is the baseline architecture and we augment a standard systolic array-based NPU with a Gathering Unit (GU); colored.
  The GU executes Feather Gathering ($\cG$) and the MAC array executes Feature Computation ($\cF$).
  GPU executes the rest, e.g., Ray Indexing ($\cI$) and the first three steps in \algo for the target frames.
  }
  \label{fig:arch}
\end{figure}

\begin{figure*}[t]
\centering
\includegraphics[width=2\columnwidth]{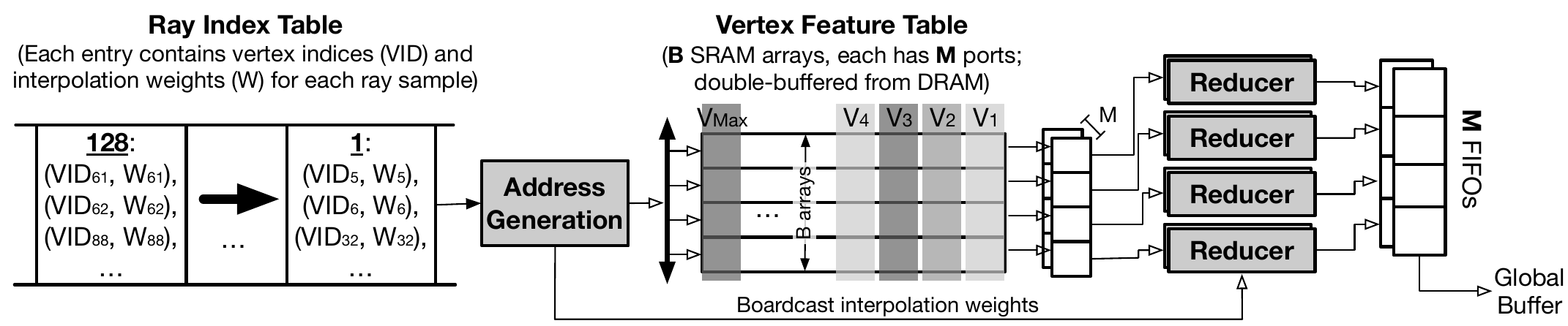}
\caption{Gathering unit.
Each RIT entry stores the information of a ray sample.
In particular, each entry has the eight vertex IDs (VIDs) of a ray sample's voxel as well as the corresponding weights (W) for trilinear interpolation.
The Address Generation logic uses the VIDs to access the corresponding vertex features from the VFT, which stores the features of a MVoxel using the channel-major layout.
The VFT has $B$ individual SRAM arrays, each has $M$ ports, supporting retrieving features for $M$ ray samples simultaneously.
$B \times M$ number of reducers perform trilinear interpolation required by the Feather Gathering ($\cG$) stage to calculate the final feature vector for $M$ ray samples.
The weights are broadcast to the reducers.
A FIFO holds the interpolated results before writing to the Global Feature Buffer for subsequent MLP computation.}
\vspace{-5pt}
\label{fig:hw_support}
\end{figure*}

\subsection{Hardware Support}
\label{sec:mem:hw}


We extend a standard NPU architecture to support the two memory optimizations.
\Fig{fig:arch} shows the SoC architecture, in which we augment the NPU with the new Gathering Unit (GU).
The baseline SoC consists of mainly a GPU and an NPU.
The GU in the NPU executes Feather Gathering ($\cG$) and the MAC array in the NPU executes Feature Computation ($\cF$).
GPU executes the rest of the computations, e.g., Ray Indexing ($\cI$) and the first three steps in \algo for the target frames.

We describe the detailed GU architecture in \Fig{fig:hw_support}.
The GU has a dedicated buffer to store the RIT, and a dedicated Vertex Feature Table (VFT) to store MVoxels streamed from the DRAM as described in \Sect{sec:mem:fs}.
The MVoxels are stored in the VFT using the data layout described in \Sect{sec:mem:bank}.
Once RIT entries are loaded to RIT, the Address Generation logic takes one RIT entry and computes the addresses of the vertex features corresponding to the ray sample in that RIT entry.

Since VFT is free from bank conflicts, the VFT is designed as $B$ individual SRAM arrays \textit{without} the area-heavy crossbar.
All the channels of the same feature vector are accessed simultaneously; thus, it takes one cycle to read one vertex feature.
Since there are eight vertices of a voxel, it takes 8 cycles to read all the feature vectors of a given ray sample.
Each bank is equipped with $M$ ports to allow $M$ feature vectors to be read simultaneously, which in turn allows $M$ ray samples to be processed in parallel.

To calculate the feature vectors for feature computation, the GU uses $B \times M$ reducers, each of which performs the trilinear interpolation operation required by NeRF models to calculate the final feature vector of a ray sample.
The interpolated feature vectors are stored in the Feature FIFO before being written to the global feature buffer in the NPU.

\paragraph{SoC Integration.}
Our hardware extensions are limited to the NPU without changing the GPU hardware.
Our design is agnostic to and, thus, integrates well with different GPU architectures --- for two reasons.  First, our hardware augmentation, i.e., the Gathering Unit, is limited to the NPU, whose communication with the GPU is dealt with by standard SoC-level interconnect (e.g., AXI) and thus accommodates different GPUs.  Second, the interaction between the NPU and the GPU is minimal in our design: the GPU simply sends the Ray Index Table through the DMA to the NPU.

\hl{\mbox{\paragraph{Broad Applicability.}}
There is a long history of co-opting graphics hardware for non-graphics workloads, starting from using shader cores for general-purpose parallel computing~\mbox{\cite{owens2007survey}} to using recent ray tracing hardware for neighbor search~\mbox{\cite{zhu2022rtnn, nagarajan2023rt}} and database operations~\mbox{\cite{nagarajan2023rtdb}}.
Our GU has the potential to be used beyond NeRF.
In essence, the GU performs a parallel gather operation followed by a parallel reduction on the gathered data.
Thus, the GU can find uses in domains such as neighbor search~\mbox{\cite{flann, pcloctreeex}} and sparse linear algebra~\mbox{\cite{duff2002overview, davis2019algorithm}}.
To accommodate different gather-reduction operations in other domains, the GU could be made programmable to allow 1) the RIT to store indices for to-be-gathered elements, 2) the address generation unit to map an element index to the element's address, 3) the VFT to store the gathered data, and 4) the Reducer to implement different reduction operations required in different applications.}

\section{Experimental Setup}
\label{sec:exp}

\paragraph{Hardware Details.}
The NPU is a systolic-array-based DNN accelerator, which has a $24 \times 24$ MAC array, where each MAC unit mimics the design of that in the TPU~\cite{jouppi2017datacenter}.
The NPU also consists of a scalar unit, which can parallelize element-wise updates such as ReLU and Max-Pooling operation.
The Global Feature Buffer is configured to be double-buffered with a size of 1.5~MB at a granularity of 32~KB.
We reserve a dedicated 96~KB weight buffer to store MLP weights.

In our GU design, the RIT is double-buffered, each sized at 6~KB to store 128 entries.
Each entry is 48 Bytes to accommodate eight vertex indices and their associated weights for linear interpolation (4-byte for one vertex index and 2-byte for one weight value).
The Vertex Feature Buffer is 32 KB \hl{(organized as $B=32$ banks each with $M=2$ ports)}, which can store a MVoxel ($8\times8\times8$ points) with 32 channels.
When the channel size of a vertex feature is greater than 32, we partition the vertex features into segments along the channel direction and load each segment sequentially.

\paragraph{Experimental Methodology.}
We directly time the GPU execution as well as the kernel launch on the mobile Volta GPU on Nvidia's Xavier SoC~\cite{xaviersoc}. The GPU power is directly measured using the built-in power sensing circuitry.
We synthesize, place, and route the datapath of the systolic array and the gathering unit using an EDA flow consisting of Synopsys and Cadence tools with the TSMC 16 nm FinFET technology and scale the results to 12 nm using the DeepScaleTool~\cite{stillmaker2017scaling, sarangi2021deepscaletool} so that the results can be comparable with the mobile Volta GPU on Nvidia's Xavier SoC in 12 nm node~\cite{xaviersoc}.

The SRAMs are generated using the Arm Artisan memory compiler. Power is estimated using Synopsys PrimeTimePX with annotated switching activities.
The DRAM is modeled after Micron 16 Gb LPDDR3-1600 (4 channels) according to its datasheet~\cite{micronlpddr3}. The DRAM energy is obtained using Micron System Power Calculators~\cite{microdrampower}. On average, the energy ratio between a random DRAM access and a streaming DRAM access is about 3:1, and the energy ratio between a random DRAM access and an SRAM access is about 25:1.

We build a cycle-level simulator of the architecture with the latency of each component parameterized from measurements (for GPU) and post-synthesis results of the NPU design.

\paragraph{Area Overhead.} \proj introduces minimal area overhead with GU augmentation. The major overhead is from 44~K SRAM introduced from RIT buffer and VFT buffer. The additional area overhead (0.048~mm$^2$) compared to baseline NPU is less than 2.5\%, in which the \proj-specific portion is almost negligible compared to the entire SoC area, such as 350 mm$^2$ for Nvidia Xavier~\cite{xaviersochotchips} and 108 mm$^2$ for Apple A15~\cite{applea15}. We also removed the crossbar connections in VFT buffer due to our interleaving access pattern in feature gathering. In comparison, a heavily banked SRAM with a crossbar would introduce an additional 0.036~mm$^2$ of area overhead.

\paragraph{NeRF Algorithms.} \proj can accommodate arbitrary NeRF algorithms. To demonstrate the flexibility of \proj, we evaluate three different NeRF algorithms: \textsc{Instant-NGP}~\cite{muller2022instant}, \textsc{DirectVoxGO}~\cite{sun2022direct} and \textsc{TensoRF}~\cite{chen2022tensorf}, \hl{with varying model size-computation trade-offs (\mbox{\Fig{fig:compute_vs_model_size}}).
The three networks also cover different feature representations: voxel grid for DirectVoxGO, hierarchical hashmaps for Instance-NGP, and factorized tensor for TensoRF.
}
We evaluate algorithm quality on a standard metric, PSNR.

\paragraph{Datasets.}
We use Synthetic-NeRF~\cite{mildenhall2021nerf}, a synthetic dataset with eight different scenes. 
We also use two real-world datasets, Unbounded-360~\mbox{\cite{barron2022mipnerf360} (Bonsai trace)} and Tanks and Temples~\mbox{\cite{Knapitsch2017}} (Ignatius trace) to evaluate \mbox{\algo} in real-world scenarios.
Generating a mesh from images in the wild is a mature field (photogrammetry).
We use Agisoft Metashape~\mbox{\cite{metashape}}, a well-known photogrammetry tool, to generate meshes for the two real-world datasets.


\paragraph{Baseline.}
Our baseline is the SoC in \Fig{fig:arch} without the GU.
It executes Ray Indexing ($\cI$) and Feature Gathering ($\cG$) on GPU and Feature Computation ($\cF$) on NPU for all frames.


\paragraph{Variants.}
We evaluate three variants of \proj to decouple the contribution proposed in our paper:
\begin{itemize}
    \item \mode{\algo}: only performs sparse radiance warping with the same hardware configuration as the baseline.
    \item \mode{\algo+FS}: same as \mode{\algo} except it includes the fully-streaming NeRF rendering.
    \item \mode{\proj}: the full version of \proj, which includes sparse radiance warping, fully-streaming NeRF rendering, and bank conflict-free interleaving (with GU support).
\end{itemize}

\paragraph{Application Scenarios.} We evaluate two application scenarios that commonly exist in AR/VR applications:
\begin{itemize}
    \item \textbf{Local Rendering}: All the computations are executed on the standalone device with the hardware described above.
    \item \textbf{Remote Rendering}: \hl{Many VR devices, such as the Oculus Quest series, can be tethered wirelessly to a remote machine (e.g., a nearby workstation or even the cloud) to accelerate rendering, whereas the local device is used for display and lightweight processing.
    How to effectively leverage the remote rendering paradigm is an active area of research, and our evaluation aims to demonstrate a particular use of remote rendering by offloading the reference frame rendering in our SPARW algorithm to a remote 2080Ti GPU via a wireless connection.}
    The wireless communication energy is modeled as 100~nJ/B with a speed of 10 MB/s~\cite{liu2022augmented}.
\end{itemize}
\section{Evaluation}
\label{sec:eval}

We first demonstrate that \proj achieves quality levels comparable to the baseline (\Sect{sec:eval:acc}). Meanwhile, we show that even a pure software implementation of \proj delivers significant speedups and energy reductions on a mobile Volta GPU (\Sect{sec:eval:gpu}).
The speedup and energy reduction increase when the baseline hardware incorporates a dedicated DNN accelerator (\Sect{sec:eval:perf}). We next perform a sensitivity study to understand \proj's performance and energy savings under different settings (\Sect{sec:eval:sen}). We show that \proj achieves better speedups compared to prior NeRF accelerators (\Sect{sec:eval:comp}).
Finally, we discuss the effectiveness of \proj on real-world datasets (\Sect{sec:eval:disc}.)
Only results in \Sect{sec:eval:disc} make use of the warping heuristics discussed in \Sect{sec:algo:dd}.

\subsection{Rendering Quality}
\label{sec:eval:acc}

\begin{figure}[t]
\centering
\subfloat[Quality evaluation on Synthetic-NeRF dataset.]{
	\label{fig:synthetic_acc}	\includegraphics[width=\columnwidth]{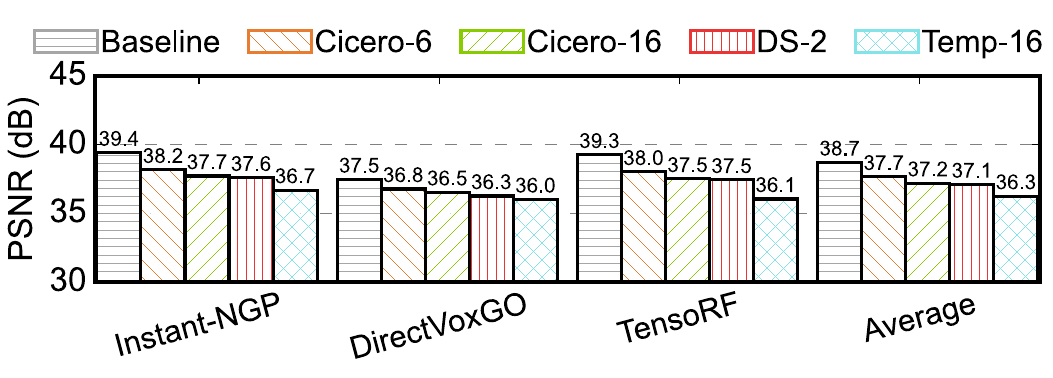} }
\\
\vspace{-10pt}
\subfloat[\hl{Quality evaluation on real-world datasets.}]{
	\label{fig:real_scene_acc}
	\includegraphics[width=\columnwidth]{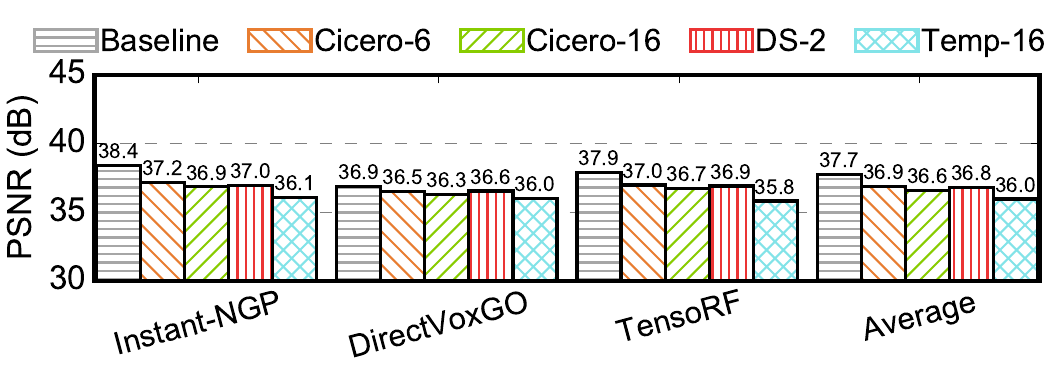} } 
\caption{Image quality comparison.
\mode{\proj-6} and \mode{\proj-16} use a warping window size (i.e., the number of target frames a reference frame is used for) of 6 and 16, respectively.
}
\label{fig:acc}
\end{figure}

\Fig{fig:acc} shows the rendering quality of applying our \algo algorithm to NeRF algorithms on both Synthetic-NeRF dataset (\Fig{fig:synthetic_acc}) and \hl{real-world scenes} (\Fig{fig:real_scene_acc}). 
We use \textit{warping window} to denote the number of target frames that reuse a single reference frame. 
We consider two warping window sizes, 6 and 16. 
In addition to the baseline algorithms, we also compare against two variants, \mode{DS-2} and \mode{Temp-16}.
\mode{DS-2} first downsamples the frame by 2 for NeRF rendering and then upsamples it to the original resolution via bilinear interpolation.
\mode{Temp-16} is a method that uses a previously rendered frame as a reference frame with a warping window of 16 frames.

On both datasets, \mode{\proj-6} retains an average PSNR drop within 1.0~dB compared to the original algorithms.
Despite \mode{\proj-16} dropping the average quality by \hl{1.3~dB}, it still has better quality compared against \mode{DS-2} and \mode{Temp-16} on Synthetic-NeRF dataset.
\mode{Temp-16} is the worst because it warps from previous frames and accumulates errors.
\hl{The quality of \mbox{\mode{\proj-6}} is only slightly better than \mbox{\mode{DS-2}} on the real-world datasets, which use a low temporal resolution (1 FPS), for which the radiance approximation does not hold well.
We will further discuss this in \mbox{\Sect{sec:eval:disc}}.}



\begin{figure}[t]
\centering
\begin{minipage}[t]{0.48\columnwidth}
  \centering
  \includegraphics[width=\columnwidth]{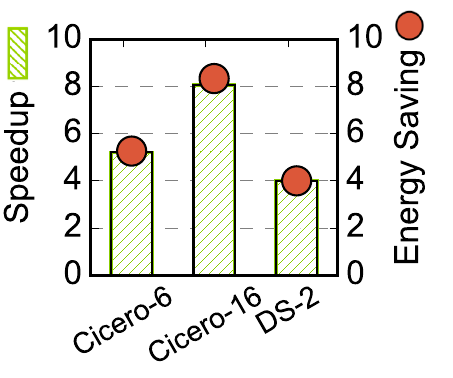}
  \caption{Speedup and energy savings of \proj against \mode{DS-2}. Numbers are normalized by GPU baselines.}
  \label{fig:gpu_result}
\end{minipage}
\hspace{2pt}
\begin{minipage}[t]{0.48\columnwidth}
  \centering
  \includegraphics[width=\columnwidth]{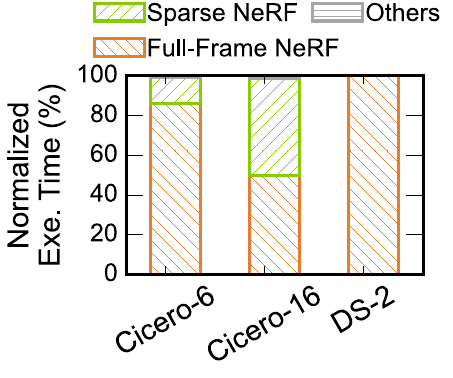}
  \caption{GPU execution distribution of \proj and \mode{DS-2}. \textit{Full-Frame NeRF} in \proj is amortized across frames.}
  \label{fig:gpu_exe_dist}
\end{minipage}
\end{figure}

\subsection{Results on GPU}
\label{sec:eval:gpu}

This section evaluates the performance of a pure software implementation of \proj.
\Fig{fig:gpu_result} compares the FPS of \mode{\proj-6} and \mode{\proj-6} with \mode{DS-2} and the results are normalized to the FPS of a mobile Volta GPU.
On average, \mode{\proj-16} achieves $8.0\times$ speedup and 7.9$\times$ energy saving compared to the original algorithms. In comparison, the \mode{DS-2} achieves only $4.0\times$ speedup and $4.0\times$ energy reduction. Even with a warping window of 6, \mode{\proj-6} is faster than \mode{DS-2}.

\Fig{fig:gpu_exe_dist} shows the execution time distribution of both \proj and \mode{DS-2}.
86.1\% of execution time in \mode{\proj-6} is due to reference frame rendering.
As the warping window size increases to 16, the percentage of Full-Frame NeRF decreases to 49.7\%,
while the execution time of sparse NeRF increases to 48.9\%.
Overall, the major latency bottleneck even with \algo is still NeRF rendering, not the warping operations in \algo.
The ``Others'' category includes all the non-NeRF operations in \algo, which is negligible.

\subsection{Performance and Energy}
\label{sec:eval:perf}

The speedup and energy reduction are even higher when the baseline SoC uses a dedicated NPU to execute the MLPs.
To demonstrate that, we evaluate two different application scenarios: local rendering vs. remote rendering, both are common in VR.
Unless stated otherwise, we use a warping window of 16 for our evaluation.


\begin{figure}[t]
\centering
\subfloat[Results of local rendering.]{
	\label{fig:standalone_res}	\includegraphics[width=\columnwidth]{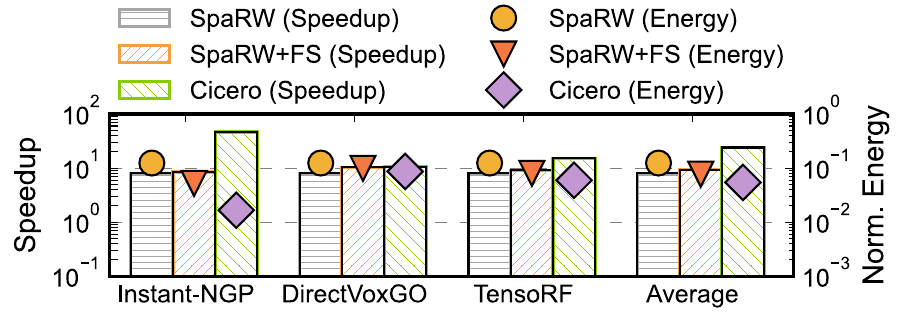} }
\\
\vspace{-5pt}
\subfloat[Results of remote rendering.]{
	\label{fig:wireless_res}
	\includegraphics[width=\columnwidth]{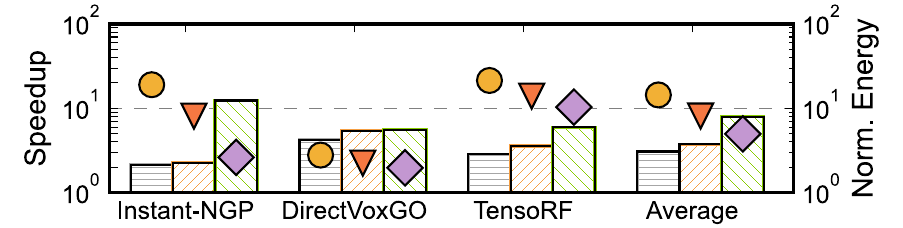} } 
\caption{End-to-end speedup and normalized energy of our variants over the baseline with a GPU and an NPU. We evaluate two application scenarios: local rendering and remote rendering. All values are normalized to the baseline.}
\label{fig:hw_speedup_energy}
\end{figure}

\paragraph{Local Rendering.} \Fig{fig:standalone_res} shows the speedup and normalized energy comparison in a local rendering scenario.
All results are normalized with the baseline.
On average, \mode{\algo} achieves 8.1$\times$ speedup and 8.1$\times$ energy saving on the same hardware configuration as the baseline.
With the additional assistance from fully-streaming NeRF rendering, \mode{\algo+FS} achieves an additional 1.2$\times$ speedup and 1.6$\times$ energy saving under the same hardware configuration. 

Two factors help \mode{\algo+FS} improve upon the baseline.
First, \algo algorithm reduces the amount of full-frame NeRF computation.
Second, fully-streaming NeRF rendering reduces redundant DRAM accesses. 
With GU hardware support, \mode{\proj} further boosts the speedup and energy saving to 28.2$\times$ and 37.8$\times$, respectively. 

\paragraph{Remote Rendering.}
One factor that prevents \proj from achieving higher speedup is resource contention: even though algorithmically reference frame and target frame rendering can be overlapped, as described in \Fig{fig:pipeline}, they compete for the same NPU and GPU resources.
With additional resources on a remote machine, \proj further boosts the performance. 

\Fig{fig:wireless_res} shows the speedup and energy comparison.
In the baseline, the entire NeRF rendering executes on the remote GPU.
In our system, we map the reference frame NeRF rendering to the remote GPU and render the target frames locally.
In both cases, the remote GPU and the local device communicate the pixel data of the rendered frames.

\mode{\algo} achieves a 3.1$\times$ speedup against the baseline, while \mode{\algo+FS} achieves a 3.8$\times$ speedup by applying fully-streaming NeRF rendering.
With GU hardware support, \mode{\proj} further improves the speedup to 8.0$\times$.
In all cases, data communication between the remote GPU and the local device is not a bottleneck: the communication latency is $0.02\%$ of the average frame latency in \proj.

Notably, the baseline in this scenario consumes lower energy than all three variants of \proj.
This is because when all the computations are offloaded to the remote GPU in the baseline, the main energy consumption of the local device is simply wireless communication.
Transferring one frame consumes almost 5$\times$ lower energy than rendering a frame in \proj.

\begin{figure}[t]
\centering
\begin{minipage}[t]{0.48\columnwidth}
  \centering
  \includegraphics[width=\columnwidth]{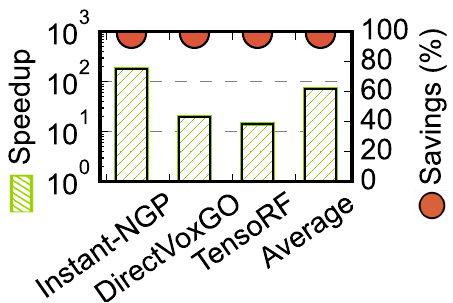}
  \caption{Speedup and energy savings of feature gathering.}
  \label{fig:gu_result}
\end{minipage}
\hspace{2pt}
\begin{minipage}[t]{0.48\columnwidth}
  \centering
  \includegraphics[width=\columnwidth]{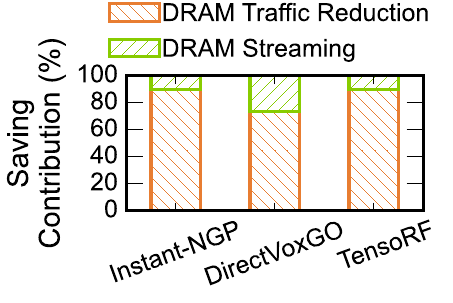}
  \caption{Memory energy saving contribution.}
  \label{fig:mem_saving_dist}
\end{minipage}
\end{figure}


\paragraph{Feature Gathering ($\cG$).}
\Fig{fig:gu_result} demonstrates the speedup and energy reduction brought by GU compared to the GPU execution. 
Overall, our GU achieves 72.2$\times$ speedup while contributing to 99.9\% of the energy reduction.
This is attributed not only to our hardware acceleration of the Gather stage, but also to our data placement strategy that eliminates bank conflicts.
For instance, Instant-NGP uses hash tables which causes severe SRAM bank conflicts. \proj eliminates the irregular accesses entirely. Coupled with GU, we achieve a 182.4$\times$ speedup on Instant-NGP.

\paragraph{Memory Saving Contribution.}
Not only does \proj eliminate non-streaming DRAM access, it also reduces the overall DRAM traffic. \Fig{fig:mem_saving_dist} plots the percentage of DRAM energy reduction attributed to DRAM traffic reduction and converting random DRAM accesses to streaming accesses.
On average, 84.5\% of energy reduction is from DRAM traffic reduction.
This shows that by grouping/loading a cluster of voxels together, \proj effectively improves the reuse of each voxel, thus reducing the overall DRAM access. The rest of the energy reduction (15.5\%) is from converting non-streaming DRAM access to streaming DRAM access. Although reducing bank conflicts does not reduce overall energy consumption, it does improve the performance of feature gathering (\Fig{fig:gu_result}).

\subsection{Sensitivity Study}
\label{sec:eval:sen}

\begin{figure}[t]
  \centering
  \captionsetup[subfigure]{width=0.5\columnwidth}
  \subfloat[\small{Local rendering.}]
  {
  \includegraphics[width=.48\columnwidth]{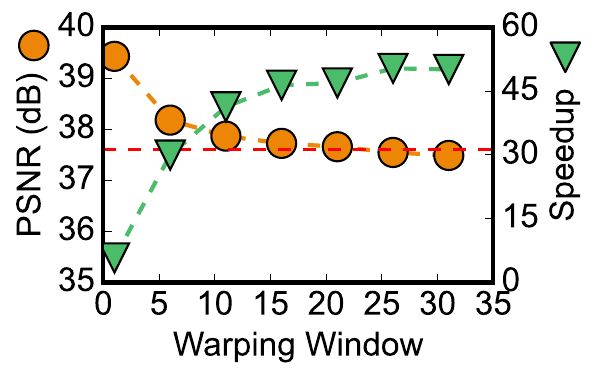}
  \label{fig:acc_sens_on_soc}
  }
  \subfloat[\small{Remote rendering.}]
  {
  \includegraphics[width=.48\columnwidth]{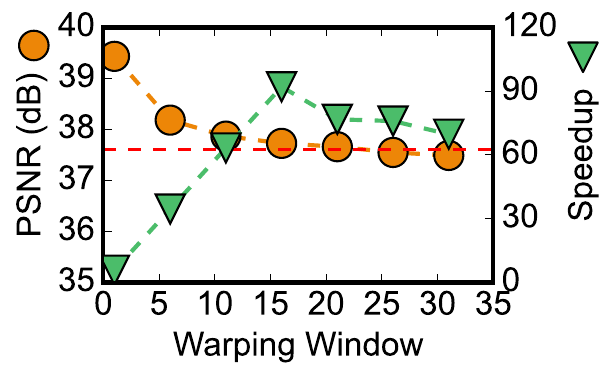}
  \label{fig:acc_sens_offload}
  }
  \caption{Sensitivity of speedup and quality on Instant-NGP to warping window size under the two scenarios. The speedup is normalized to the baseline under the local rendering scenario. The red dash line shows the quality of \mode{DS-2}.}
  \label{fig:acc_sens}
\end{figure}

\paragraph{Warping Window.} \Fig{fig:acc_sens} illustrates how the overall speedup and quality vary under different warping window lengths (energy follows a similar trend as speed-up).
For simplicity, we focus on Instant-NGP.

\Fig{fig:acc_sens_on_soc} shows the sensitivity to the warping window length in the local rendering scenario. Unsurprisingly, the quality of \proj gradually decreases as the warping window increases. Nevertheless, \proj still retains higher quality compared to \mode{DS-2} in \Fig{fig:acc}, even at a warping window of 21.
Similarly, the speedup of \proj gradually plateaus and starts to decline as the warping window reaches 26.
The decline is due to the growing number of empty pixels caused by disocclusions (\Fig{fig:intuition}).
Consequently, the workload of sparse NeRF rendering, which aims to fill the disocclusions, gradually increases and eventually becomes dominant. 

In comparison, \Fig{fig:acc_sens_offload} shows the sensitivity in the remote rendering scenario.
One interesting observation is that the speedup of \proj first increases linearly as the warping window increases from 1 to 16. This is because the on-device execution can be hidden behind the execution time of full-time NeRF rendering offloaded to the workstation.
However, when the warping window reaches 16, the on-device execution time can no longer be hidden and becomes the dominant factor.

\begin{figure}[t]
\centering
\begin{minipage}[t]{0.48\columnwidth}
  \centering
  \includegraphics[width=\columnwidth]{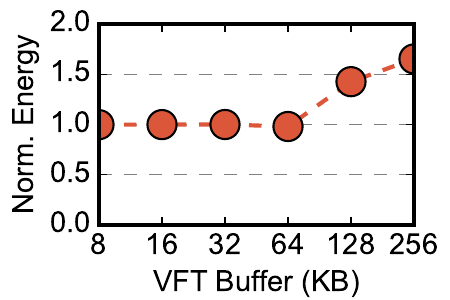}
  \caption{Sensitivity of GU energy consumption to different VFT buffer sizes.}
  \label{fig:gu_sens}
\end{minipage}
\hspace{2pt}
\begin{minipage}[t]{0.48\columnwidth}
  \centering
  \includegraphics[width=\columnwidth]{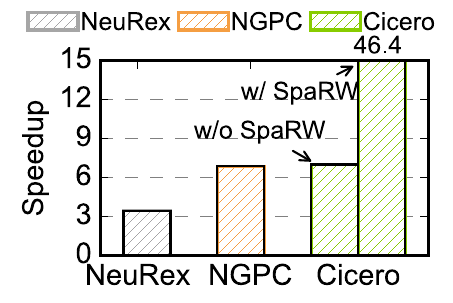}
  \caption{Comparison of speedup between \proj and prior NeRF accelerators.}
  \label{fig:prior_work_comp}
\end{minipage}
\end{figure}

\paragraph{GU.} \Fig{fig:gu_sens} shows the sensitivity of GU energy consumption to various VFT buffer sizes. To ensure each bank can hold one channel value of a MVoxel, we maintain the bank size at 1KB while sweeping the VFT buffer sizes (effectively bank sizes). To accommodate the same number of reducers in GU, we increase the number of ports per bank. As shown in \Fig{fig:gu_sens}, the overall GU energy consumption remains relatively constant as VFT buffer size increases from 8~KB to 64~KB. However, beyond 64~KB, the overall GU energy starts to rise.

\subsection{NGPC and NeuRex Comparison}
\label{sec:eval:comp}


We also compare against two prior NeRF accelerators: \textsc{NeuRex}~\cite{lee2023neurex} and \textsc{NGPC}~\cite{mubarik2023hardware}.
Notably, both two accelerators are tailored to one particular NeRF algorithm, Instant-NGP, whereas \proj generally applies to any NeRF algorithms.

\hl{\mbox{\Fig{fig:prior_work_comp}} compares \mbox{\proj} with the two accelerators on Instant-NGP.
All values are normalized to the GPU baseline.
We use the data reported in the \textsc{NeuRex} paper\footnote{The original NeuRex paper compares against Xavier NX (21 TOPS, 384 core) and our GPU baseline is Xavier (32 TOPS, 512 core).
To convert the result to the Xavier baseline, we use the actual execution time of Instant-NGP on Xavier NX and NeuRex’s speedup numbers reported in the original paper to calculate the absolute execution time of NeuRex.
Based on that, we calculate the speed-up of NeuRex over Xavier used in \mbox{\Fig{fig:prior_work_comp}}.} and implement the \textsc{NGPC} architecture based on the paper's description.
For a fair comparison, we configure \mbox{\proj} to use the same amount of PEs ($24 \times 24$) as NGPC; NeuRex has a higher PE count ($32 \times 32$) as reported in the paper.
Our accelerator uses a 32 KB feature buffer, and the two baseline accelerators have larger on-chip feature buffers as described in their respective papers (i.e. 16 MB for NGPC and 64 KB for NeuRex).}


Overall, \proj without \algo algorithm demonstrates a 2.0$\times$ speedup over \textsc{NeuRex}.
The speedup against \textsc{NeuRex} is attributed to hardware augmentation of GU in \proj, which eliminates the SRAM bank conflicts in feature gathering.
By contrast, \textsc{NGPC} design inherently avoids SRAM bank conflicts (because they use one bank for all the feature vectors in one Instant-NGP level).
\proj without \algo achieves a similar speed.
However, \textsc{NGPC} requires a 16~MB on-chip buffer dedicated to storing feature encodings, which is unrealistic for a mobile SoC.
In contrast, with fully-streaming rendering algorithm, our on-chip SRAM size is only 32 KB.
With our \algo algorithm, \proj boosts the speedup to 16.4$\times$ and 8.2$\times$ against \textsc{NeuRex} and \textsc{NGPC}, respectively.

\subsection{Discussion on Real-World Scenes}
\label{sec:eval:disc}

\begin{figure}[t]
  \centering
  \captionsetup[subfigure]{width=0.5\columnwidth}
  \subfloat[\small{Sparse (1 FPS) version.}]
  {
  \includegraphics[width=.48\columnwidth]{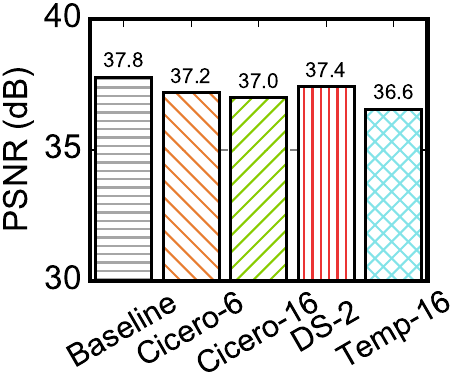}
  \label{fig:sparse_acc}
  }
  \subfloat[\small{Dense (30 FPS) version.}]
  {
  \includegraphics[width=.48\columnwidth]{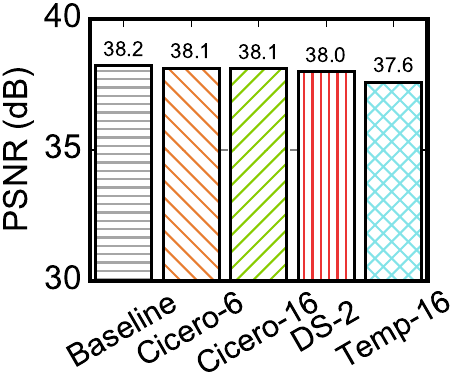}
  \label{fig:dense_acc}
  }
  \caption{\hl{PSNR comparison on the \textit{Ignatius} scene in the Tanks and Temples dataset. The temporally dense sequence is more representative of real-time VR rendering, which we target.}}
  \label{fig:sparse_dense_acc}
\end{figure}

\hl{We use the \textit{Ignatius} scene in the Tanks and Temples dataset to discuss the effectiveness and limitations of \mbox{\algo} on real-world scenes. 
\mbox{\Fig{fig:sparse_acc}} shows the results.
Both \mbox{\mode{\proj-6}} and \mbox{\mode{\proj-16}} have lower quality compared to \mbox{\mode{DS-2}}.
This is because the temporal resolution of the scene is extremely low (1 FPS).
Thus, consecutive frames have large differences in camera poses (i.e., $\theta$ in \mbox{\Fig{fig:intuition}} is too large), so the radiance approximation does not hold well for non-diffuse surface.

We hasten to stress that the lower quality of \mbox{\algo} here is \textit{not} fundamental to the algorithm but an artifact of the low-FPS dataset.
To evaluate \mbox{\proj} in scenarios more representative of real-time VR rendering, we use the raw video sequence from the dataset captured at 30 FPS.
The results are shown in \mbox{\Fig{fig:dense_acc}}.
In this more realistic scenario, \mbox{\mode{\proj-16}} has little quality loss over the baseline and has a similar quality compared to \mbox{\mode{DS-2}} but is about 4$\times$ faster.}

\begin{figure}
    \centering
    \includegraphics[width=\columnwidth]{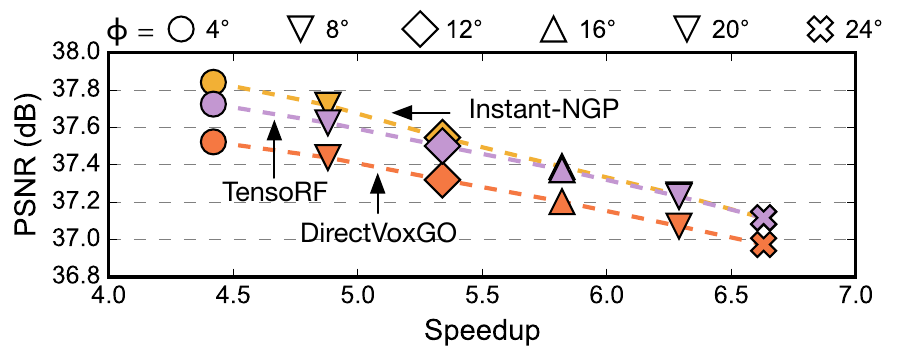}
    \caption{\hl{Speed-up and PSNR of \mbox{\mode{\proj-16}} under different warping thresholds $\phi$ on the 1-FPS sequence.  We forgo warping if the angle subtended by a ray in the reference frame and the corresponding ray in the target frame is greater than $\phi$}.}
    \label{fig:angle_threshold}
\end{figure}

While real-time VR rendering, which we target, usually has a high temporal resolution ($>$ 30 FPS), in scenarios where a dataset has low temporal resolution, our warping heuristics in \Sect{sec:algo:dd} can be used to mitigate the rendering quality loss.
\mbox{\Fig{fig:angle_threshold}} shows the speed-up and PSNR of \mbox{\mode{\proj-16}} across different warping thresholds on the challenging 1-FPS sequence of \textit{Ignatius}.
As $\phi$ reduces toward the left, the quality increases, since fewer pixels are warped and more pixels are NeRF-rendered, which also means the performance reduces.
At a threshold of 4$^{\circ}$, \mbox{\algo} has a quality drop within 0.1 dB and a speed-up of 4.3$\times$.

\section{Related Work}
\label{sec:related}

\paragraph{NeRF Acceleration.} NeRF rendering has drawn considerable attention in the last two years. Recent works have proposed several accelerators for NeRF algorithms~\cite{lee2023neurex, rao2022icarus, li2023instant, mubarik2023hardware, li2022rt, fu2023gen}.
However, prior designs are tailored to individual NeRF algorithms—one accelerator for one algorithm.
For example, Instant-3D~\cite{li2023instant} and \textsc{NeuRex}~\cite{lee2023neurex} accelerate the training and inference of Instant-NGP~\cite{muller2022instant}, respectively. NGPC~\cite{mubarik2023hardware} accelerates a range of neural graphic algorithms with similar hierarchical feature encodings.
ICARUS~\cite{rao2022icarus} and RT-NeRF~\cite{li2022rt}, on the other hand, design specialized architecture to speed up NeRF~\cite{mildenhall2021nerf} and TensoRF~\cite{chen2022tensorf}, respectively. Meanwhile, Gen-NeRF~\cite{fu2023gen} accelerates IBRNet~\cite{wang2021ibrnet} for novel view synthesis.  
In contrast, \proj proposes solutions that are generally applicable to a range of existing NeRF algorithms.

\hl{\mbox{\paragraph{Memory Optimizations in Rendering.}}
Our idea of full-streaming DRAM accesses is inspired by ray reordering techniques in conventional ray tracing that increase memory access locality by grouping nearby rays~\mbox{\cite{pharr1997rendering, shkurko2017dual,aila2010architecture,bikker2012improving,gribble2008coherent}}.

Our approach has three main differences.
First, we change the basic unit of reordering from rays to ray samples to accommodate the nature of NeRF.
Second, existing techniques usually manipulate rays \textit{dynamically}, since (secondary) rays are spawned at run time, which complicates the hardware design (e.g., dynamic identification of nearby rays, dynamic buffer management).
In contrast, we exploit the nature of NeRF where all ray samples are known statically and reorder ray samples only once at the beginning.
Third, ray reordering in ray tracing usually can only afford local reordering because rays are dynamically spawned, whereas we reorder ray samples \textit{globally} to guarantee fully-streaming DRAM accesses.}

\no{
\paragraph{Real-Time VR Rendering.} To achieve real-time VR rendering, prior works have leaned on image-based rendering or remote rendering~\cite{teler2001streaming, boos2016flashback, mueller2018shading, hladky2022quadstream, han2023metavrain, leng2019energy}.
Some systems directly stream rendered videos to clients, but they often suffer from bandwidth limits and require efficient video encoding~\cite{noimark2003streaming}. Other works transmit one frame that can be reused to render multiple viewpoints, but often require complex encoding methods to store texture and geometric information to address disocclusions~\cite{mueller2018shading, hladky2022quadstream}.
By leveraging the computation characteristics of NeRF, \proj proposes a straightforward yet effective approach to resolve disocclusions, achieving photo-realistic visual quality.
}

\paragraph{Warping in Image-based Rendering.}
\algo belongs to a class of imaging warping techniques initially proposed for image-based rendering~\cite{chen2023view, chen1995quicktime}, of which NeRF is a recent development.
We refer interested readers to Chaurasia et al.~\cite{chaurasia2020passthrough+} and Szeliski~\cite{szeliski2022image} for brief surveys.
Recent studies generalize warping to use temporal correlations across frames for reducing computation in real-time vision~\cite{buckler2018eva2, zhu2018euphrates, feng2019asv, ying2022exploiting, song2020vr, hou2023architecting}, increase super-resolution~\cite{xiao2020neural} and depth estimation quality~\cite{feng2023fast}, and accelerate rendering~\cite{zhao2020deja, zhao2021holoar}. 

We are the first to apply temporal warping in NeRF (as a way of view synthesis).
Prior temporal warping techniques serialize the processing of the reference frame and the target frame, as the latter depends on the former.
\mbox{\algo}'s main novelty is the observation that the reference frames merely provide useful information for target frames later.
We break the dependencies between reference and target frames, which allows us to overlap reference and target frame rendering.



\section{Limitations and Future Work}
\label{sec:lim}

\paragraph{Warping.}
The warping heuristics discussed in \Sect{sec:algo:dd} is nothing more than an engineering hack to accommodate non-diffuse surfaces.
Ideally, we want a ``transfer function'' that transforms the radiance of a ray to the radiance of another ray.
The \algo essentially uses an identity function (conditioned upon the warping threshold) as a special-case transfer function.

The exact transfer function depends on the material property of the surface.
An interesting future direction is to learn the material properties~\cite{gkioulekas2013inverse, azinovic2019inverse}, e.g, Bidirectional Reflectance Distribution Function (BRDF) and Bidirectional Subsurface Scattering Function Function (BSSDF)~\cite{pharr2023physically}, potentially jointly with the NeRF model.
The material property could then be used to better transfer pixels in the reference frame, similar to classic irradiance caching~\cite{krivanek2022practical, ward1988ray} and radiance caching~\cite{muller2021real, krivanek2005radiance, seyb2020design}, instead of simply reusing the pixel values.


\algo also exposes potential aliasing issues across the boundary between warped pixels and NeRF-rendered pixels.
One simple solution would be blended across the regions using techniques from classic foveated rendering~\cite{patney2016towards, guenter2012foveated}.

\paragraph{3DGS.} Similar to MLP-based models, 3DGS follows the pixel-centric approach during image rendering. Thus, our ideas in principle can also be applied to 3DGS. Furthermore, 3DGS avoids MLP computations and uses point cloud-based representations, which might lead to higher memory consumption~\cite{feng2020mesorasi} and irregular memory access~\cite{feng2022crescent} during Feature Gathering, warranting future investigations.
\section{Conclusion}
\label{sec:conc}

We reduce over 95\% of the MLP computation in NeRF by warping radiances computed in previous frames with less than 1 dB PSNR loss.
We also show that transforming NeRF inference from a ray-centric order to a scene-centric order leads to a completely sequential DRAM access.
Finally, we show that laying feature vectors in a channel-major, rather than a feature-major, order eliminates on-chip SRAM bank conflicts.
Collectively, we demonstrate over an order of magnitude speed-up and energy saving over a mobile Volta GPU.

\section{Acknowledgements}

We thank anonymous reviewers from ISCA for their comments and Weikai Lin from University of Rochester for invaluable discussion and technical support. Jingwen Leng is the corresponding author. The work is partially supported by the National Key R\&D Program of China under Grant 2022YFB4501400, NSFC grant (62072297 and 62222210).


\bibliographystyle{IEEEtranS}
\interlinepenalty=10000
\bibliography{refs}

\end{document}